\documentclass{aastex62}

\usepackage{amsmath}
\usepackage{color}
\usepackage{booktabs}
\usepackage{multirow}
\usepackage{lineno}

\received{2022 April 15}
\revised{2022 August 10}
\accepted{2022 August 11}

\shorttitle{A supercritical accretion disk with outflows}
\shortauthors{Cao \& Gu}

\begin{document}
\title{A supercritical accretion disk with radiation-driven outflows}

\author{Xinwu Cao}
\affiliation{Institute for Astronomy, School of Physics, Zhejiang University, 866 Yuhangtang Rd, Hangzhou, 310058, China, Email: xwcao@zju.edu.cn}

\author{Wei-Min Gu}
\affiliation{Department of Astronomy, Xiamen University, Xiamen, Fujian 361005, China}




\begin{abstract}
Outflows are inevitably driven from the disk if the vertical component of the black hole (BH) gravity cannot resist the radiation force. We derive the mass loss rate in the outflows by solving a dynamical equation for the vertical gas motion in the disk. The structure of a supercritical accretion disk is calculated with the radial energy advection included. We find that most inflowing gas is driven into outflows if the disk is accreting at a moderate Eddington-scaled rate (up to $\sim 100$) at its outer edge, i.e., only a small fraction of gas is accreted by the BH, which is radiating at several Eddington luminosities, while it reaches around ten for extremely high accretion rate cases ($\dot{m}\equiv\dot{M}/\dot{M}_{\rm Edd}\sim 1000$). Compared with a normal slim disk, the disk luminosity is substantially suppressed due to the mass loss in the outflows. We apply the model to the light curves of the tidal disruption events (TDEs), and find that the disk luminosity declines very slowly with time even if a typical accretion rate $\dot{m}\propto t^{-5/3}$ is assumed at the outer edge of the disk, which is qualitatively consistent with the observed light curves in some TDEs, and helps to understanding the energy deficient phenomenon observed in the TDEs. Strong outflows from supercritical accretion disks surrounding super-massive BHs may play crucial roles on their host galaxies, which can be taken as an ingredient in the mechanical feedback models. The implications of the results on the growth of supper-massive BHs are also discussed.
\end{abstract}

\keywords{accretion, accretion disks -- black hole physics -- ISM: jets and outflows.}

\section{Introduction}\label{intro}

Black hole (BH) accretion is the main power source of active galactic nuclei (AGNs), X-ray binaries, and tidal disruption events (TDEs), the observed features of which can be well reproduced by the thin accretion disk model  \citep*[e.g.,][]{1973A&A....24..337S,1989MNRAS.238..897L,2016ApJ...821..104Y,2018MNRAS.474.3593K}, if their Eddington-scaled luminosity is not very high (less than $\sim 0.3$)\citep*[e.g.,][]{1989MNRAS.238..897L}. The observations show that a fraction of AGNs are luminous, which are radiating at super-Eddington luminosities \citep*[e.g.,][]{2002ApJ...579..530W,2006ApJ...648..128K,2009MNRAS.398.1905W,2012ApJ...746..169S,2013ApJ...764...45K,2014ApJ...787L..20D,2021ApJ...920....9L}. For those luminous sources, the thin disk assumption is no longer satisfied, and a slim disk model was developed by \citet{1988ApJ...332..646A}, which have been extensively studied in the past decades \citep*[e.g.,][]{1988ApJ...330..142E,1988ApJ...332..646A,1996ApJ...458..474S,1999ApJ...522..839W,
2000PASJ...52..133W,2001ApJ...552L.109K,2003A&A...398..927W,2003ApJ...593...69K,2007ApJ...660..541G,2010ApJ...715..623L,2011A&A...527A..17S,2022arXiv220403606W}. In the slim disk model, the radiation pressure is assumed to be balanced with the vertical component of the BH gravity, and the scale-height of the disk is around unity due to its strong radiation from the disk. Its viscous timescale is shorter than the cooling timescale, and therefore the radial energy advection is important compared with the viscously dissipated energy, which reduces the radiation power of the slim disk. The slim disk luminosity increases with the accretion rate in a non-linear way, i.e., its radiation efficiency decreases with increasing accretion rate \citep*[][]{1988ApJ...332..646A}. The radiation of a slim disk can be substantially higher than the Eddington luminosity if the accretion rate is sufficiently high \citep*[e.g.,][]{2003ApJ...593...69K,2005ApJ...628..368O,2006ApJ...648..523W}.

{The vertical hydrostatic equilibrium assumption in the disk indicates a maximal radiation flux, above which there is no hydrostatic disk solution.} It implies that the gas may inevitably be driven away from the disk surface when the disk radiation is strong with a relative half disk thickness $H/R>\sqrt{2}/2$ \citep*{2006ApJ...652..518M,2007ApJ...660..541G,2015MNRAS.448.3514C}. The radiation hydrodynamical simulations indeed show that strong winds are always associated with luminous disks, and in some cases most gas inflowing onto the disk is driven into the outflows, while only a small fraction of the gas is finally swallowed by the BH \citep*[e.g.,][]{2005ApJ...628..368O,2011ApJ...736....2O,2014MNRAS.441.3177M,2014ApJ...780...79Y,2014ApJ...796..106J,
2019ApJ...880...67J,2021PASJ...73..450K,2022arXiv220314994H}. The outflows are
observationally discovered in the high soft states of black hole (BH) X-ray binaries \citep*[e.g.,][]{2016ApJ...821..104Y}. Observations also show that strong outflows are associated with luminous quasars \citep*[e.g.,][]{2003ApJ...593L..11Y,2003MNRAS.346.1025P,2013ApJ...764..150M,2017MNRAS.471..706J,2017MNRAS.472L..15M,2018ApJ...865....5R,2019ApJ...878...79L,2020ApJ...896..151R,2021ApJ...920....9L}. A theoretical model has been developed for the structure and radiation of such a luminous disk with radiation driven outflows, which is found significantly different from the that of the slim disk without outflows \citep*[][]{2015MNRAS.448.3514C,2019ApJ...885...93F}. In their model, the mass loss rate in the outflows is estimated on the assumption that it can automatically adjust the radiation flux to a critical value to maintain a vertical hydrostatic state of the disk with $H/R=\sqrt{2}/2$. This assumption may not be good enough for the reality, because the disk should not be in strictly vertical hydrostatic state in the presence of outflows, at least in the upper layer near the disk surface, where the gas is leaving from the disk into the outflows.

In this work, we investigate the vertical structure of a supercritical accretion disk with outflows by solving the dynamical equation of the disk in the vertical direction, from which the mass loss rate in the outflows is derived. The radial structure of a disk with outflows is then calculated, in which both the mass loss rate in the outflows and the radial energy advection have been properly included. We describe the model in Section \ref{model}, and the results in Section \ref{results}. The final section contains the discussion of the results.

\vskip 1cm
\section{Model}\label{model}

It is usually assumed that the vertical gravitational force is in balance with the radiation force in the luminous thin disks. However, the vertical component of gravity of the BH increases along $z$, and reaches the maximum at
$z/R=\sqrt{2}/2$, which implies the gas above $z/R=\sqrt{2}/2$ will be blown away if the radiation of the disk surpasses a critical value  \citep*[][]{2006ApJ...652..518M,2007ApJ...660..541G}. In our previous works, the structure of a disk with outflows has been derived by assuming the outflow to be strong to carry away sufficient gas from the disk surface and the disk radiation is limited at a critical value \citep*[][]{2015MNRAS.448.3514C,2019ApJ...885...93F}. However, such an assumption may not always be satisfied, for example, if a large amount of gas is inflowing onto the outer region of the disk, the radiation-driven outflows may not be able to carry sufficient gas away from the disk, which may lead to the radiation flux surpassing the critical value derived on the assumption of vertical hydrostatic equilibrium. In this work, we solve the dynamical equation of the vertical gas motion in the disk, and the mass loss rate in the outflows from the unit area of the disk surface is derived. With the mass loss in the outflows, we develop a self-consistent supercritical accretion disk model with outflows, in which the radial energy advection in the disk and the mass loss and kinetic power of the outflows have been properly considered.

\subsection{Vertical structure of the disk}\label{z_struct}

We first consider a radiation-pressure-dominated disk accreting at a moderate rate without radiation-driven outflows. The vertical hydrostatic equilibrium of the accretion disk requires
\begin{equation}
{\frac {{d}p(z)}{{d}z}}=-{\frac {GM\rho z}{(R^2+z^2)^{3/2}}},
\label{vert_disc1}
\end{equation}
at radius $R$, {where $G$ is the gravitational constant, $M$ is the BH mass, $p$ is the pressure, and $\rho$ is the density of the gas in the disk}. The radiation transfer in the vertical direction is described by
\begin{equation}
f_{\rm rad}(z)=-{\frac {4\sigma}{3\kappa_{\rm T}\rho(z)}}{\frac {\partial T^4(z)}{\partial z}}=-{\frac {4\sigma}{3}}{\frac {\partial T^4}{\partial \tau}},\label{rad_transfer}
\end{equation}
where $\kappa_{\rm T}$ is the electron scattering opacity, {$\sigma$ is the Stefan-Bolzmann constant, $f_{\rm rad}$ is the radiation flux in the vertical direction, and $T$ is the temperature of the gas in the disk}. The optical depth $\tau(z)$ is defined as
\begin{equation}
\tau(z)=\int\limits_0^z \rho(z^\prime)\kappa_{\rm T}dz^\prime. \label{tau}
\end{equation}
Using Equation (\ref{rad_transfer}), we obtain
\begin{equation}
{\frac {dp(z)}{dz}}=-{\frac {\kappa_{\rm T}\rho(z)f_{\rm rad}(z)}c},
\label{p_f_rad}
\end{equation}
where {$c$ is the light speed}, and $p=aT^4/3$ {($a$ is the radiation constant)} is used for a radiation pressure dominated accretion disk. Substituting Equation (\ref{p_f_rad}) into Equation (\ref{vert_disc1}), we have
\begin{equation}
{\frac
{GMz}{(R^2+z^2)^{3/2}}}={\frac {f_{\rm rad}(z)\kappa_{\rm T}}{c}}.
\label{dedz1}
\end{equation}

The vertical radiation flux $f_{\rm rad}(z)$ increases with $z$, and it is contributed by the gravitational power released in the disk. We further assume that the power added into the radiation flux is proportional to the gas density in the disk, i.e.,
\begin{equation}
{\frac {df_{\rm rad}(z)}{dz}}={\frac {\rho(z)}{\bar{\rho}H}}f_{\rm rad}(H),  \label{f_rad_z}
\end{equation}
or
\begin{equation}
f_{\rm rad}[\tau(z)]={\frac {\tau(z)}{\tau(H)}}f_{\rm rad}(H),  \label{f_rad_z2}
\end{equation}
where $\bar{\rho}$ is the mean density of the disk, $H$ is the half-thickness of the disk, and $f_{\rm rad}(H)$ is the outgoing flux at the disk surface, which is similar to that assumed in \citet{1973A&A....24..337S} for a standard thin accretion disk. Substituting Equation (\ref{dedz1}) into (\ref{f_rad_z}), we obtain the vertical density distribution of the disk,
\begin{equation}
\rho(z)={\frac {R^2-2z^2}{(R^2+z^2)^{5/2}}}(R^2+H^2)^{3/2}\bar{\rho}.\label{rho_z_nw}
\end{equation}
The outgoing radiation flux from the unit surface area of the
disk at $z=H$ is available,
\begin{equation}
f_{\rm rad}(H)={\frac {GMH}{(R^2+H^2)^{3/2}}}{\frac
{c}{\kappa_{\rm T}}}, \label{f_rad1}
\end{equation}
with Equation (\ref{dedz1}). {Inspecting the left part of Equation (\ref{dedz1}), one may easily find that the vertical BH gravity increases with $z$ from the mid-plane of the disk until  $z=\sqrt{2}R/2$, and then it decreases with $z$.} Accordingly, the radiation flux $f_{\rm rad}$ reaches a critical value when $H=\sqrt{2}R/2$ \citep*[][]{2006ApJ...652..518M,2007ApJ...660..541G}, which means that, the maximal half disk thickness of a hydrostatic disk $H_{\rm max}\equiv\sqrt{2}R/2$. In the case of $f_{\rm rad}>f_{\rm rad}^{\rm crit}$, the gas at the disk surface will be inevitably driven away by the radiation force of the disk, because the vertical gravitational force decreases with $z$ when $z>\sqrt{2}R/2$ and it is unable to resist the radiation force \citep*[][]{2015MNRAS.448.3514C}. Therefore, substituting $H=\sqrt{2}R/2$ into Equation (\ref{f_rad1}), we have
\begin{equation}
f_{\rm rad}^{\rm crit}={\frac {2\sqrt{3}GMc}{9R^2\kappa_{\rm T}}},
 \label{f_rad_crit}
\end{equation}
above which the radiation force will drive the disk gas into outflows
\citep*[see][for the details]{2015MNRAS.448.3514C}. {We assume that the disk is always in vertical hydrostatic equilibrium whenever there is a solution to Equation (\ref{vert_disc1}), i.e.,  $f_{\rm rad}\le f_{\rm rad}^{\rm crit}$, which has been widely adopted in the previous works \citep*[e.g.,][]{1973A&A....24..337S,1988ApJ...332..646A}. In this case, the half disk thickness $H$ can be calculated with Equation (\ref{f_rad1}) if $f_{\rm rad}(H)$ is specified, and then the vertical disk structure is available with Equation (\ref{rho_z_nw}).}

In the case of $f_{\rm rad}>f_{\rm rad}^{\rm crit}$, the vertical hydrostatic equilibrium is never satisfied in the disk, i.e., no solution to the hydrostatic equilibrium disk equation (\ref{vert_disc1}) is available. The vertical motion of the gas is therefore described by
\begin{equation}
\rho(z)v_z(z){\frac {dv_z(z)}{dz}}=-{\frac {{d}p(z)}{{d}z}}-{\frac {GM\rho(z) z}{(R^2+z^2)^{3/2}}},
\label{vert_disc2}
\end{equation}
{where $v_z$ is the vertical velocity of the gas in the disk}. Substituting Equation (\ref{p_f_rad}) into (\ref{vert_disc2}), the vertical momentum equation becomes
\begin{equation}
v_z(z){\frac {dv_z(z)}{dz}}={\frac {\kappa_{\rm T}f_{\rm rad}(z)} {c}}-{\frac {GMz}{(R^2+z^2)^{3/2}}}.
\label{vert_disc3}
\end{equation}

Combining Equations (\ref{rad_transfer}) and (\ref{f_rad_z2}), we have
\begin{equation}
-{\frac {4\sigma}{3}}{\frac {dT^4}{d\tau}}={\frac {\tau(z)}{\tau(H)}}f_{\rm rad}(H).\label{dtdz}
\end{equation}
A solution to this equation is available,
\begin{equation}
T^4[\tau(z)]=T_{\rm s}^4-{\frac {3f_{\rm rad}(H)}{8\sigma\tau(H)}}\tau^2(z)+{\frac {3f_{\rm rad}(H)}{8\sigma}}\tau(H),\label{t_tau}
\end{equation}
with suitable boundary conditions at $z=0$ and $H$, where $T_{\rm s}=T(H)$ is the surface temperature of the disk. The gas temperature at the mid-plane of the disk,
\begin{equation}
T_{\rm c}^4=T_{\rm s}^4+{\frac 3{8\sigma}}f_{\rm rad}(H)\tau(H)\simeq {\frac 3{8\sigma}}f_{\rm rad}(H)\tau(H),\label{t_c}
\end{equation}
for an optically thick accretion disk [i.e., $\tau(H)\gg 1$], as $f_{\rm rad}(H)=\sigma T_{\rm s}^4$. The vertical pressure distribution of the gas in a radiation pressure dominant accretion disk is available,
\begin{equation}
p[\tau(z)]={\frac {f_{\rm rad}(H)}{c}}\left[{\frac 4 3}+{\frac {\tau(H)}{2}}-{\frac {\tau^2(z)}{2\tau(H)}}\right],
\label{p_tau}
\end{equation}
with Equation (\ref{t_tau}), as $p=aT^4/3$ is assumed for a radiation pressure dominated disk. {In order to derive the vertical density distribution of the disk, we need a relation of the density with pressure. In principle, the equation of state, $p=aT^4/3+\rho kT/\mu m_{\rm p}$ ($k$ is the Boltzmann's constant, and $\mu$ is the atomic weight of the plasma), acts as such a relation, though the radiation pressure always dominates over the gas pressure in the supercritical accretion disk. To avoid the complexity of numerical calculations of the vertical structure of the disk, instead, we adopt a polytropic relation, $p=K\rho^\gamma$ ($\gamma$ is the polytropic index), in deriving the vertical structure of the disk, as done in many previous works \citep*[e.g.,][]{1981AcA....31..283P,1988ApJ...332..646A,2002apa..book.....F,2008bhad.book.....K}. Thus, we derive the density distribution as
\begin{equation}
\rho(\tau)=\left[1+{\frac 3 8}\tau(H)-{\frac {3\tau^2(z)}{8\tau(H)}}   \right]^{1/\gamma}\rho(H)
\label{rho_tau}
\end{equation}
from Equation (\ref{p_tau}).}

We derive the relation of $\tau(z)$ with $z$,
\begin{displaymath}
z=\int\limits_0^\tau {\frac {d\tau^\prime}{\kappa_{\rm T}\rho(\tau^\prime)}}={\frac 1{\kappa_{\rm T}\rho(H)}}\int\limits_0^\tau \left[1+{\frac 3 8}\tau(H)-{\frac {3\tau^{\prime 2}(z)}{8\tau(H)}}\right]^{-1/\gamma}d\tau^\prime
\end{displaymath}
\begin{equation}
={\frac {H\bar{\rho}}{\tau(H)\rho(H)}}
\int\limits_0^\tau \left[1+{\frac 3 8}\tau(H)-{\frac {3\tau^{\prime 2}(z)}{8\tau(H)}}\right]^{-1/\gamma}d\tau^\prime,
\label{tau_z}
\end{equation}
with
\begin{equation}
{\frac {d\tau(z)}{dz}}=\kappa_{\rm T}\rho(z),\label{dtaudz}
\end{equation}
and Equation (\ref{rho_tau}).
Letting $z=H$ in Equation (\ref{tau_z}), the gas density at the disk surface is
\begin{equation}
{\frac {\rho(H)}{\bar{\rho}}}={\frac 1{\tau(H)}}\int\limits_0^{\tau(H)}\left[1+{\frac 3 8}\tau(H)-{\frac {3\tau^{\prime 2}(z)}{8\tau(H)}}\right]^{-1/\gamma}d\tau^\prime.
\label{rho_h}
\end{equation}
{In the region with $f_{\rm rad}(H)>f_{\rm rad}^{\rm crit}$, we note that the interior gas in the disk flows outwards in the vertical direction, and it is subtle to define the thickness of the disk. As afore discussion of the thickness of a vertically hydrostatic equilibrium disk, its maximal thickness $H_{\rm max}=\sqrt{2}R/2$ corresponding to $f_{\rm rad}(H)=f_{\rm rad}^{\rm crit}$. In the region with $z>H_{\rm max}$, the $z$-component of the BH gravity is not able to resist the radiation force. We therefore assume that the half disk thickness $H\equiv H_{\rm max}\equiv \sqrt{2}R/2$ in the disk region with $f_{\rm rad}(H)>f_{\rm rad}^{\rm crit}$ where the outflows are leaving the disk.}

Now we are able to obtain the relation of $z$ with $\tau$ with Equations (\ref{tau_z}) and (\ref{rho_h}), and then the vertical density distribution $\rho(z)$ of the disk with  Equation (\ref{rho_tau}), when the values of the half-thickness $H$, and the optical depth $\tau(H)$ of the disk are specified. Finally, the vertical motion of the gas in the disk is derived by integrating Equations (\ref{f_rad_z2}) and (\ref{vert_disc3}) for a given outgoing energy flux $f_{\rm rad}(H)>f_{\rm rad}^{\rm crit}$. With derived dynamics of the vertical motion of the gas in the disk, the mass loss rate of the outflows from the unit area of the disk surface is available,
\begin{equation}
\dot{m}_{\rm w}=\rho(H)v_z(H). \label{mdot_w}
\end{equation}


\subsection{Radial structure of the disk}\label{r_struct}

In the presence of outflows, the continuity equation of the disk is
\begin{equation}
{\frac {{d}\dot{M}(R)}{{d}R}}=4\pi R\dot{m}_{\rm w}(R),
\label{mdot_acc}
\end{equation}
where $\dot{M}$ is the accretion rate of the disk at $R$, and $\dot{m}_{\rm w}$ is the mass loss rate in the outflow from the unit area of the disk surface. Otherwise, $d\dot{M}(R)/dR=0$ for a normal accretion disk without outflows.

For a supercritical accretion disk, the power tapped into the outflows  and the radial  advection power  play important roles, so we include these two terms in the energy equation of the disk {all these quantities are for one single surface of the disk}, and hereafter, we use $f_{\rm rad}=f_{\rm rad}(H)$, and $\tau=\tau(H)$ for convenience],
\begin{equation}
f_{\rm rad}=Q^+-Q_{\rm adv}-Q_{\rm w},
\label{energy}
\end{equation}
where
\begin{equation}
Q_{\rm w}={\frac 1 2}\dot{m}_{\rm w}v_z^2(H)={\frac 1 2}\rho(H)v_z^3(H),
\label{q_w}
\end{equation}
\begin{equation}
Q_{\rm adv}={\frac {\dot{M}}{4\pi R^2}}{\frac {p_{\rm c}}{\rho_{\rm c}}}\xi_{\rm adv},\label{q_adv}
\end{equation}
and
\begin{equation}
\xi_{\rm adv}=4{\frac {{d\ln}\rho_{\rm c}}{{d\ln}R}}-12{\frac {{d\ln}T_{\rm c}}{{d\ln}R}},\label{xi_adv}
\end{equation}
is adopted for a radiation pressure dominant accretion disk \citep*[][]{1995ApJ...438L..37A,1999ApJ...516..420W,2000ApJ...540L..33G}. {As only the kinetic power carried away by the outflow from the disk appears in Equation (\ref{q_w}), one may wonder about the contribution from the internal energy of the gas driven into the outflow. In order to clarify this issue, we integrate the term due to the internal energy in the general form of the energy equation for the gas over $z$ \citep*[see, Equation 2.5 in][]{2002apa..book.....F},
\begin{equation}
\int\limits_0^H \nabla\cdot(\rho\epsilon \boldsymbol{v})dz={\frac 1 R}{\frac {\partial}{\partial R}}(R\bar{\rho}H\epsilon v_R)+\dot{m}_{\rm w}\epsilon={\frac \epsilon R}{\frac {\partial}{\partial R}}(R\bar{\rho}H v_R)+\bar{\rho}Hv_R{\frac {\partial \epsilon}{\partial R}}+\dot{m}_{\rm w}\epsilon,
\label{div_e}
\end{equation}
for an axis-symmetric disk, where $v_R$ is the radial velocity, and $\epsilon$ is the internal energy per unit mass. Substituting the continuity equation (\ref{mdot_acc}) into Equation (\ref{div_e}), we have
\begin{equation}
\int\limits_0^H \nabla\cdot(\rho\epsilon \boldsymbol{v})dz=\bar{\rho}Hv_R{\frac {\partial \epsilon}{\partial R}},
\label{div_e2}
\end{equation}
where $\dot{M}=-4\pi R \bar{\rho}Hv_R$ is used. We note that the term $\dot{m}_{\rm w}\epsilon$ in Equation (\ref{div_e}) is canceled out, which implies that the effect of the internal energy of the gas driven into the outflows on the disk has been described by the continuity equation of the disk implicitly.}

{The angular momentum equation for a steady accretion disk is
\begin{equation}
{\frac {\partial}{\partial R}}\left(R\nu\Sigma R^2{\frac {d\Omega}{dR}}\right)=-{\frac 1{2\pi}}{\frac d{dR}}(\dot{M}R^2\Omega)+{\frac 1{2\pi}}R^2\Omega{\frac {d\dot{M}}{dR}},\label{angular}
\end{equation}
where $\nu$ is the viscosity, $\Sigma\equiv 2H\bar{\rho}$ is the surface density of the disk, and $\Omega$ is the angular velocity of the gas in the disk \citep*[see,][]{2002apa..book.....F}. {In this work, the angular velocity $\Omega$ can be understood as the vertically averaged value, and the angular momentum transfer in the vertical direction of the disk has not been considered (see Sect. \ref{z_struct}). The angular velocity of the gas leaving from the disk surface may deviate from $\Omega$, which would contribute a bit to the angular momentum equation (\ref{angular}). However, we believe it may not deviate much from the vertically averaged $\Omega$. Therefore, to avoid the complexity of the vertical angular momentum transfer in the disk, we assume its contribution to the angular momentum equation of the disk to be negligible. We note that the hydrodynamical outflows are in stark contrast with the magnetically driven outflows, in which the gas may move along the field lines co-rotating with the disk to a distance far from the disk surface, and therefore a substantial fraction of the disk angular momentum may be removed through the magnetically driven outflows \citep*[e.g.,][]{1994MNRAS.268.1010L,2002A&A...385..289C,2013ApJ...765..149C}.}

Integrating Equation (\ref{angular}) over $R$, we obtain
\begin{equation}
\nu\Sigma R^3{\frac {d\Omega}{dR}}=-{\frac {\dot{M}(R)}{2\pi}}R^2\Omega+{\frac {\dot{M}(R_{\rm in})}{2\pi}}R_{\rm in}^2\Omega(R_{\rm in})+{\frac {L_{\rm w}(R)}{2\pi}}-{\frac {L_{\rm w}(R_{\rm in})}{2\pi}}, \label{angular_2}
\end{equation}
by adopting the boundary condition, $d\Omega/dR=0$ at the inner radius of the disk $R_{\rm in}$ \citep*[][]{1973A&A....24..337S}, and the relation $-2\pi R\Sigma(R)v_R=\dot{M}(R)$ is used. {The rate of angular momentum $L_{\rm w}(R)$ carried away by the outflows from the disk region between $R$ and $R_{\rm out}$ is
\begin{equation}
L_{\rm w}=\int\limits_{R}^{R_{\rm out}}{R^\prime}^2\Omega{\frac {d\dot{M}}{dR^\prime}}dR^\prime=4\pi\int\limits_{R}^{R_{\rm out}}\dot{m}_{\rm w}(R^\prime){R^\prime}^2\Omega R^\prime dR^\prime, \label{l_w}
\end{equation}
where Equation (\ref{mdot_acc}) is used.}

The viscously dissipated power in the disk is
\begin{equation}
Q^+={1 \over 2}\nu\Sigma\left(R{\frac {d\Omega}{dR}}\right)^2={\frac {3GM\dot{M}(R)\tilde{\Omega}^2(R)}{8\pi R^3}}f(R, \dot{M}, \tilde{\Omega}, L_{\rm w}),
\label{q_plus}
\end{equation}
where the factor,
\begin{equation}
f(R, \dot{M}, \tilde{\Omega}, L_{\rm w})=1-{\frac {\dot{M}(R_{\rm in})R_{\rm in}^{1/2}\tilde{\Omega}(R_{\rm in})}{\dot{M}(R)R^{1/2}\tilde{\Omega}(R)}}
-{\frac {L_{\rm w}(R)}{\dot{M}R^2\Omega}}+{\frac {L_{\rm w}(R_{\rm in})}{\dot{M}R^2\Omega}},\label{f_cor}
\end{equation}
is induced by the inner boundary condition of the disk, $\tilde{\Omega}=\Omega/\Omega_{\rm K}$, $\Omega_{\rm K}$ is the Keplerian rotational velocity, and the approximation $d\Omega/dR\simeq -3\Omega/2R$ is adopted.} For a non-rotating BH, $R_{\rm in}=6GM/c^2$.

The relation of the radiation flux at the disk surface with the temperature $T_{\rm c}$ at the mid-plane of an optically thick disk is (see Equation \ref{t_c})
\begin{equation}
f_{\rm rad}\simeq{\frac {8\sigma T_{\rm c}^4}{3\bar{\rho} H\kappa_{\rm T}}}={\frac {8\sigma T_{\rm c}^4}{3\rho_{\rm c}BH\kappa_{\rm T}}},
\label{T_c}
\end{equation}
where the correction factor,
\begin{equation}
B={\frac {\bar{\rho}}{\rho_{\rm c}}}=(1+\tilde{H}^2)^{-3/2},\label{b_rho_nw}
\end{equation}
in the region with $f_{\rm rad}\le f_{\rm rad}^{\rm crit}$ (see Equation \ref{rho_z_nw}), while the value of $B$ is available from the calculation of the vertical dynamics in the disk with outflows as described in Section \ref{z_struct}.

The radial velocity $v_R$ of the gas in the disk is available with the angular momentum equation (\ref{angular_2}) of the disk,
\begin{equation}
v_R(R)=-{\frac 3 2}\alpha c_{\rm s}{\frac H R}f(R, \dot{M}, \tilde{\Omega}, L_{\rm w})^{-1}
=-{\frac 3 2}\alpha \left({\frac {p_{\rm c}}{\rho_{\rm c}}}\right)^{1/2}{\frac H R}f(R, \dot{M}, \tilde{\Omega}, L_{\rm w})^{-1},
\label{v_r}
\end{equation}
where the $\alpha$-viscosity $\nu=\alpha c_{\rm s} H$ is adopted, and
\begin{equation}
{\frac {p_{\rm c}}{\rho_{\rm c}}}={\frac {H\kappa_{\rm T}B}{2c}}f_{\rm rad},
\label{p_rho}
\end{equation}
is derived with Equation (\ref{T_c}). {Strictly speaking, the sound speed $c_{\rm s}$ is also a function of the polytropic index $\gamma$. To be consistent with the most commonly used formalism in the literature, we still adopt $c_{\rm s}=(p_{\rm c}/\rho_{\rm c})^{1/2}$ in this work. This is equivalent to such inaccuracy absorbed by the viscosity parameter $\alpha$, of which the value is still quite uncertain.}

The mass accretion rate in the disk is
\begin{equation}
\dot{M}(R)=-2\pi R\Sigma v_R=-4\pi R\bar{\rho} H v_R=6\pi\alpha H^2B  p_{\rm c}^{1/2}\rho_{\rm c}^{1/2}f(R, \dot{M}, \tilde{\Omega}, L_{\rm w})^{-1}.
\label{mdot}
\end{equation}
{It is well known that the gas in a supercritical accretion disk rotates at a sub-Keplerian velocity \citep*[][]{1988ApJ...332..646A}, which is mainly caused by the radial pressure gradient. The rotational velocity of the disk is estimated with the radial momentum equation,
\begin{equation}
{\frac 1{\rho_{\rm c}}}{\frac {dp_{\rm c}}{dR}}\simeq -{\frac {p_{\rm c}}{R\rho_{\rm c}}}=R\Omega^2-R\Omega_K^2,
\label{v_phi}
\end{equation}
where the term $dv_R/dR=0$ is assumed. This is a fairly good approximation in the disk region not very close to the inner edge of the disk. Substituting Equation (\ref{p_rho}) into Equation (\ref{v_phi}), we have
\begin{equation}
\tilde{\Omega}=\left[1-{\frac {H\kappa_{\rm T}B f_{\rm rad}}{2R^2\Omega_{\rm K}^2 c}}    \right]^{1/2}.
\label{v_phi2}
\end{equation}}
Differentiating Equations (\ref{T_c}), and (\ref{mdot}), the value of $\xi_{\rm adv}$ is available with Equation (\ref{xi_adv}),
\begin{equation}
\xi_{\rm adv}=-{\frac 7 2}{\frac {d{\ln}f_{\rm rad}}{d{\ln}R}}-{\frac {11} 2}{\frac {d{\ln} H}{d{\ln}R}}+{\frac {d{\ln}f}{d{\ln}R}}+{\frac {d{\ln}\dot{M}}{d{\ln}R}}.\label{xi_adv3}
\end{equation}
Substituting Equations (\ref{q_adv}) and (\ref{xi_adv3}) into Equation (\ref{energy}), we obtain
\begin{equation}
{\frac {df_{\rm rad}}{dR}}=-{\frac {11f_{\rm rad}}{7R}}{\frac {d{\ln}H}{d{\ln}R}}
+{\frac {2f_{\rm rad}}{7R}}{\frac {d{\ln}f}{d{\ln}R}}
+{\frac 2 7}\left[{\frac {f_{\rm rad}}{R}}+{\frac {v_z^2(H)c}{RH\kappa_{\rm t}B}}\right]{\frac {d{\ln}\dot{M}}{d{\ln}R}}
-{\frac {6GMc\tilde{\Omega}^2f}{7R^2H\kappa_{\rm T}B}}
+{\frac {16\pi Rcf_{\rm rad}}{7\dot{M}H\kappa_{\rm T}B}}, \label{dfraddr2}
\end{equation}
where $d\dot{M}/dR$ is given by Equations (\ref{mdot_w}) and (\ref{mdot_acc}) in the disk region with outflows.

\vskip 0.5cm

\subsubsection{The disk region with outflows}\label{woutflow}

In the region of the disk with $f_{\rm rad}>f_{\rm rad}^{\rm crit}$, the mass loss rate of the outflows is calculated with Equation (\ref{mdot_w}) as described in Section \ref{z_struct}, and
$H=H_{\rm max}\equiv\sqrt{2}R/2$ is assumed (see the discussion in Section \ref{z_struct}). For convenience, we define the dimensionless quantities as
\begin{displaymath}
r={\frac {R}{R_{\rm g}}},~~R_{\rm g}={\frac {GM}{c^2}},~~\tilde{H}={\frac H R},~~\tilde{z}={\frac z R},~~\lambda={\frac {L}{L_{\rm Edd}}},~~L_{\rm Edd}={\frac {4\pi GMc}{\kappa_{\rm T}}},~~\dot{M}_{\rm Edd}={\frac {L_{\rm Edd}}{0.1c^2}},~~\dot{m}={\frac {\dot{M}}{\dot{M}_{\rm Edd}}},~~\tilde{f}_{\rm rad}={\frac {R_{\rm g}^2\kappa_{\rm T}f_{\rm rad}}{GMc}},
\end{displaymath}
\begin{equation}
 \tilde{Q}^+={\frac {R_{\rm g}^2\kappa_{\rm T}Q^+}{GMc}},~~\tilde{Q}_{\rm w}={\frac {R_{\rm g}^2\kappa_{\rm T}Q_{\rm w}}{GMc}},~~\tilde{L}_{\rm w}={\frac {L_{\rm w}}{\dot{M}(R_{\rm in})R_{\rm in}^2\Omega(R_{\rm in})}},~~\tilde{\rho}(z)={\frac {\rho(z)}{\bar{\rho}}},~~{\rm and}~~ \tilde{v}_z=\left({\frac R{GM}}\right)^{1/2}v_z,
\label{dimensionless}
\end{equation}
{where $R_{\rm g}$ is the gravitational radius, $L_{\rm Edd}$ is the Eddington luminosity, and $\dot{M}_{\rm Edd}$ is the Eddington accretion rate}. The dimensionless energy equation becomes
\begin{equation}
{\frac {d\tilde{f}_{\rm rad}}{dr}}=-{\frac {11\tilde{f}_{\rm rad}}{7r}}
+{\frac {2\tilde{f}_{\rm rad}}{7r}}{\frac {d{\ln} f}{d{\ln}r}}
-{\frac {6\tilde{\Omega}^2f}{7\tilde{H}r^3B}}+{\frac {2\tilde{f}_{\rm rad}}{35\tilde{H}B\dot{m}}}
+{\frac 2 7}\left[{\frac {\tilde{f}_{\rm rad}}r}+{\frac {\tilde{v}_z^2(H)}{\tilde{H}Br^3}}\right]{\frac {d{\ln}\dot{m}}{d{\ln}r}},\label{energy3}
\end{equation}
where
\begin{equation}
{\frac {d{\ln}f}{d{\ln}r}}={\frac {\dot{m}(r_{\rm in})r_{\rm in}^{1/2}\tilde{\Omega}(r_{\rm in})}{f\dot{m}(r)r^{1/2}\tilde{\Omega}(r)}}\left({\frac 1 2}+{\frac {d{\ln}\dot{m}}{d{\ln}r}}  \right)[1+\tilde{L}_{\rm w}(r)-\tilde{L}_{\rm w}(r_{\rm in})]-{\frac 1 f}{\frac {d{\ln}\dot{m}}{d{\ln}r}}.
\label{dlnfdlnr}
\end{equation}

{The rotational velocity of the gas in the disk is
\begin{equation}
\tilde{\Omega}=\left(1-{\frac 1 2}\tilde{H}Br^2\tilde{f}_{\rm rad}\right)^{1/2},
\label{v_phi3}
\end{equation}
in the dimensionless form (see Equation \ref{v_phi2}).}

The continuity equation (\ref{mdot_acc}) can be re-written in the dimensionless form,
\begin{equation}
{\frac {d\dot{m}}{dr}}=0.1\tilde{H}^{-1}_{\rm max}\tau\tilde{\rho}(H)\tilde{v}_z(H)r^{-1/2},
\label{mdot_acc2}
\end{equation}
where Equation (\ref{mdot_w}) is adopted. The optical depth of the disk is
\begin{equation}
\tau=\bar{\rho}H\kappa_{\rm T}=-{\frac {\dot{M}(R)\kappa_{\rm T}}{4\pi Rv_R(R)}}.
\label{tau}
\end{equation}
Substituting Equations (\ref{v_r}) and (\ref{p_rho}) into Equation (\ref{tau}), we obtain
\begin{equation}
\tau={\frac {20\sqrt{2}}3}\alpha^{-1}\dot{m}(r)\tilde{H}^{-3/2}r^{-3/2}B^{-1/2}\tilde{f}_{\rm rad}^{-1/2}f(r, \dot{m}, \tilde{\Omega}, \tilde{L}_{\rm w})
\label{tau_2}
\end{equation}
in the dimensionless form.
The dynamics of the vertical gas motion in the disk are calculated as described in Section \ref{z_struct}, and then the mass loss rate $\dot{m}_{\rm w}$ in the outflows is available.

\subsubsection{The disk region without outflows}\label{wooutflow}

In the region with $f_{\rm rad}\le f_{\rm rad}^{\rm crit}$, no outflow is driven from the disk surface, i.e., $\dot{m}_{\rm w}=0$, and the radiation flux $f_{\rm rad}$ is related to the disk thickness with Equation (\ref{f_rad1}). Differentiating Equations (\ref{f_rad1}), we have
\begin{equation}
{\frac {d{\ln}H}{d{\ln}R}}={\frac {R^2+H^2}{R^2-2H^2}}{\frac {d{\ln}f_{\rm rad}}{d{\ln}R}}+{\frac {3R^2}{R^2-2H^2}}.
\label{dlnhdlnr}
\end{equation}
Substituting Equation (\ref{dlnhdlnr}) into Equation (\ref{dfraddr2}), and letting $d\dot{M}/dR=0$,  we obtain
\begin{equation}
{\frac {df_{\rm rad}}{dR}}=-{\frac {11R}{6R^2-H^2}}f_{\rm rad}
+{\frac {2(R^2-2H^2)f_{\rm rad}}{3R(6R^2-H^2)}}{\frac {d{\ln}f}{d{\ln}R}}
-{\frac {2(R^2-2H^2)GMc\tilde{\Omega}^2f}{(6R^2-H^2)R^2H\kappa_{\rm T}B}}
+{\frac {16(R^2-2H^2)\pi Rcf_{\rm rad}}{3(6R^2-H^2)\dot{M}H\kappa_{\rm T}B}}.\label{dfraddr}
\end{equation}
The dimensionless energy equation for the disk region without outflows is
\begin{equation}
{\frac {d\tilde{f}_{\rm rad}}{dr}}=-{\frac {11\tilde{f}_{\rm rad}}{r(6-\tilde{H}^2)}}
+{\frac {2(1-2\tilde{H}^2)\tilde{f}_{\rm rad}}{3r(6-\tilde{H}^2)}}{\frac {d{\ln}f}{d{\ln}r}}
-{\frac {2(1-2\tilde{H}^2)\tilde{\Omega}^2f}{(6-\tilde{H}^2)\tilde{H}r^3B}}
+{\frac {2(1-2\tilde{H}^2)\tilde{f}_{\rm rad}}{15(6-\tilde{H}^2)\tilde{H}\dot{m}B}}.
\label{energy2}
\end{equation}
The vertical hydrostatic equilibrium equation (\ref{f_rad1}) can be re-written as
\begin{equation}
\tilde{f}_{\rm rad}={\frac {\tilde{H}}{r^2(1+\tilde{H}^2)^{3/2}}},\label{f_rad2}
\end{equation}
in dimensionless form for the region with $\tilde{f}_{\rm rad}\le \tilde{f}_{\rm rad}^{\rm crit}$, {with which the half disk thickness $\tilde{H}$ can be calculated for a given radiation flux $\tilde{f}_{\rm rad}$}.
The dimensionless viscously dissipated power is
\begin{equation}
\tilde{Q}^+(r)={\frac {15\dot{m}(r)\tilde{\Omega}^2(r)}{r^3}}f(r, \dot{m}, \tilde{\Omega}, \tilde{L}_{\rm w}),
\label{q_plus2}
\end{equation}
which is valid for the disk regions either with or without outflows. The Eddington ratio of the disk is
\begin{equation}
\lambda={\frac L{L_{\rm Edd}}}={\frac {\kappa_{\rm T}}{4\pi GMc}}\int\limits_{R_{\rm in}}^{R_{\rm out}}4\pi Rf_{\rm rad}{d}R=\int\limits_{r_{\rm in}}^{r_{\rm out}} r \tilde{f}_{\rm rad}dr.
\label{lum_disk}
\end{equation}

\section{Results}\label{results}

The structure of a supercritical accretion disk with radiation-driven outflows is available by integrating Equations (\ref{energy3}) or (\ref{energy2}) for the regions of the disk with or without outflows respectively,
with suitable boundary conditions. Combing Equations (\ref{energy}) and (\ref{q_adv}), we have
\begin{equation}
{\frac {\dot{M}(R)}{4\pi R^2}}{\frac p\rho}\xi_{\rm adv}=Q^+(R)-f_{\rm rad}(R)-Q_{\rm w}(R),
\label{q_adv2}
\end{equation}
which can be re-written in the dimensionless form as
\begin{equation}
\tilde{f}_{\rm rad}(r)=[(\tilde{Q}^+(r)-\tilde{Q}_{\rm w}(r)]\left[{\frac {5\dot{m}(r)\tilde{H}(r)\xi_{\rm adv}B(r)}{r}}+1\right]^{-1},
\label{q_adv3}
\end{equation}
where
\begin{equation}
\tilde{Q}_{\rm w}={\frac {\tilde{\rho}(H)\tilde{v}_z^3(H)\tau}{2\tilde{H}r^{5/2}}},
\label{q_w2}
\end{equation}
$\tilde{H}=\tilde{H}_{\rm max}\equiv\sqrt{2}/2$ if $\tilde{f}_{\rm rad}>\tilde{f}_{\rm rad}^{\rm crit}\equiv 2\sqrt{3}/9r^2$, otherwise, $\tilde{H}$ is related to $\tilde{f}_{\rm rad}$ with Equation (\ref{f_rad2}), and $\tilde{Q}_{\rm w}=0$, for the disk region without outflow, i.e., $\tilde{f}_{\rm rad}\le \tilde{f}_{\rm rad}^{\rm crit}$. In this work, we assume $\xi_{\rm adv}=3/2$ at $r=r_{\rm out}$ as suggested in a self-similar slim disk solution \citep*[][]{1999ApJ...516..420W}, and therefore we obtain the value of $\tilde{f}_{\rm rad}(r_{\rm out})$ with Equation (\ref{q_adv3}). Then we integrate Equation (\ref{energy3}) together with Equation (\ref{mdot_acc2}) in the region with $\tilde{f}_{\rm rad}>\tilde{f}_{\rm rad}^{\rm crit}$ (or Equation \ref{energy2} in the region with $\tilde{f}_{\rm rad}\le\tilde{f}_{\rm rad}^{\rm crit}$) from $r=r_{\rm out}$.
The model calculations are carried out with the specified values of three parameters, namely, the viscosity parameter $\alpha$, the accretion rate $\dot{m}(r_{\rm out})$ at the outer radius of the disk $r_{\rm out}$. The accretion rate at the inner radius of the disk $\dot{m}(r_{\rm in})$, {the rate of the angular momentum carried away by the outflows $\tilde{L}_{\rm w}(r_{\rm in})$}, and $\tilde{\Omega}(r_{\rm in})$ are not model parameters, which are fine tuned to let the solution satisfy the inner boundary conditions at $r_{\rm in}$ (so-called ``shooting method"). {We note that $\tilde{\Omega}$ is very close to the unity at $r=r_{\rm in}$, because a zero torque inner boundary condition is adopted (i.e., $d\Omega/dR=0$ at $R=R_{\rm in}$).} We find that the final results are quite insensitive to the value of $\xi_{\rm adv}$ at $r_{\rm out}$.

In the region with radiation driven outflows, i.e., $f_{\rm rad}>f_{\rm rad}^{\rm crit}$, we calculate the dynamics of the vertical motion of the gas  in the disk as described in Section \ref{z_struct}. The values of $f_{\rm rad}$ and $\dot{M}$ at a certain radius $R$ are derived with integrating the radial equations of the disk, and then the optical depth $\tau$ of the disk is available with Equation (\ref{tau_2}). The density $\rho(\tau)$ of the disk as a function of $\tau$ is given by Equation (\ref{rho_tau}). Integrating Equation (\ref{tau_z}), we obtain a relation of $z$ with $\tau$, and therefore the vertical density distribution $\rho(z)$ is available. A polytropic index $\gamma$ is adopted in the calculation of the vertical disk structure. {The detailed calculations of the vertical structure of a radiation pressure dominant disk indicates $\gamma=1.5$ \citep*[][]{1981AcA....31..283P}, which is also adopted for a slim disk in \citet{1988ApJ...332..646A}. We adopt $\gamma=1.5$ in the most model calculations.} The vertical momentum equation (\ref{vert_disc3}) can be re-written as
\begin{equation}
\tilde{v}_z(\tilde{z}){\frac {d\tilde{v}_z(\tilde{z})}{d\tilde{z}}}=r^2\tilde{f}_{\rm rad}(\tilde{z})-{\frac {\tilde{z}}{(1+\tilde{z}^2)^{3/2}}},
\label{vert_disc4}
\end{equation}
in the dimensionless form. Integrating this momentum equation, in which the flux $f_{\rm rad}(z)$ is calculated with Equation (\ref{f_rad_z}), the velocity $v_z(z)$ of the vertical motion in the disk is derived. Using Equation (\ref{mdot_w}), the mass loss rate $\dot{m}_{\rm w}$ in the disk at a certain radius is calculated. We repeated the vertical calculations at every radius, and finally obtained the whole structure of the disk with outflows.

\begin{figure}
\gridline{\fig{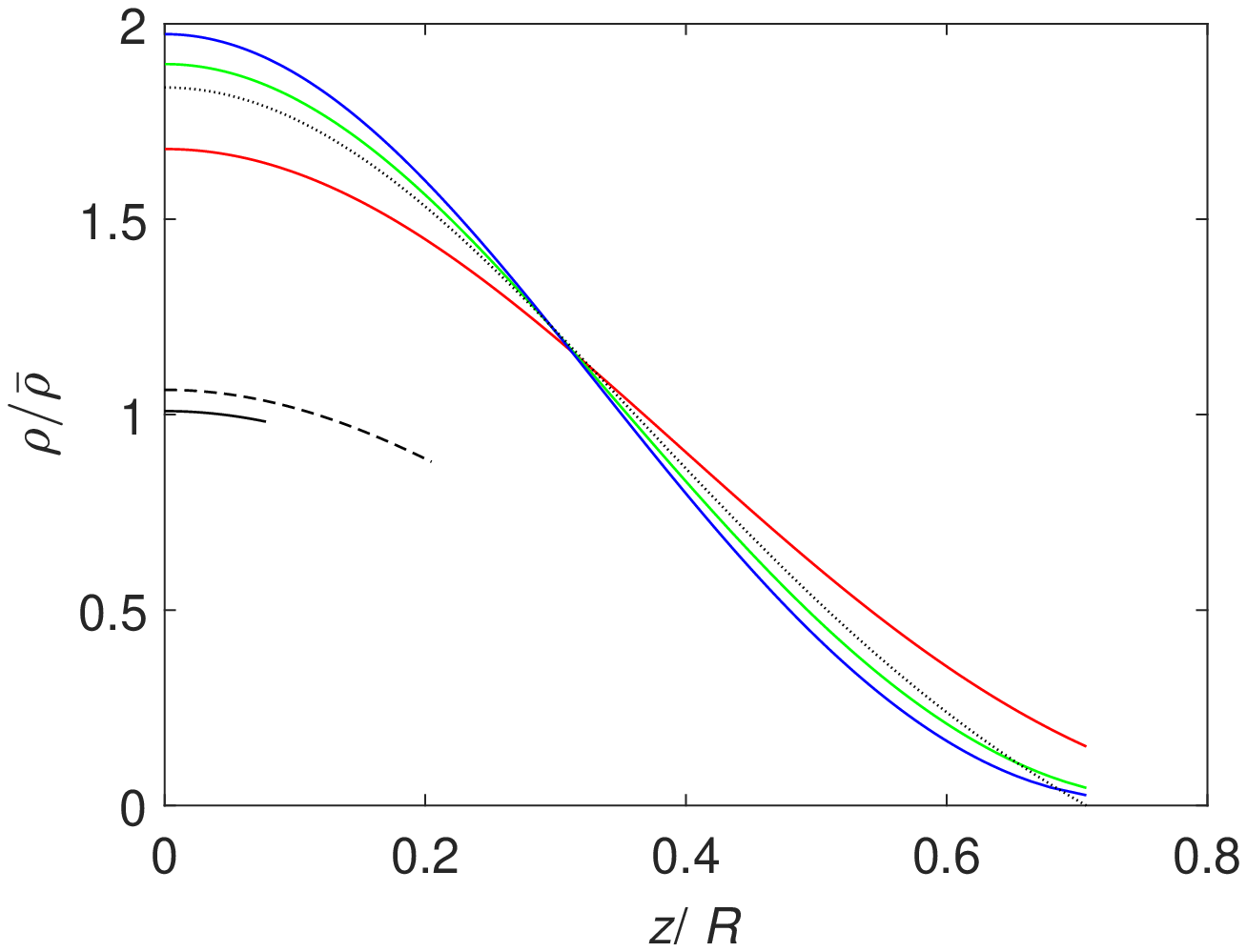}{0.4\textwidth}{}
 \fig{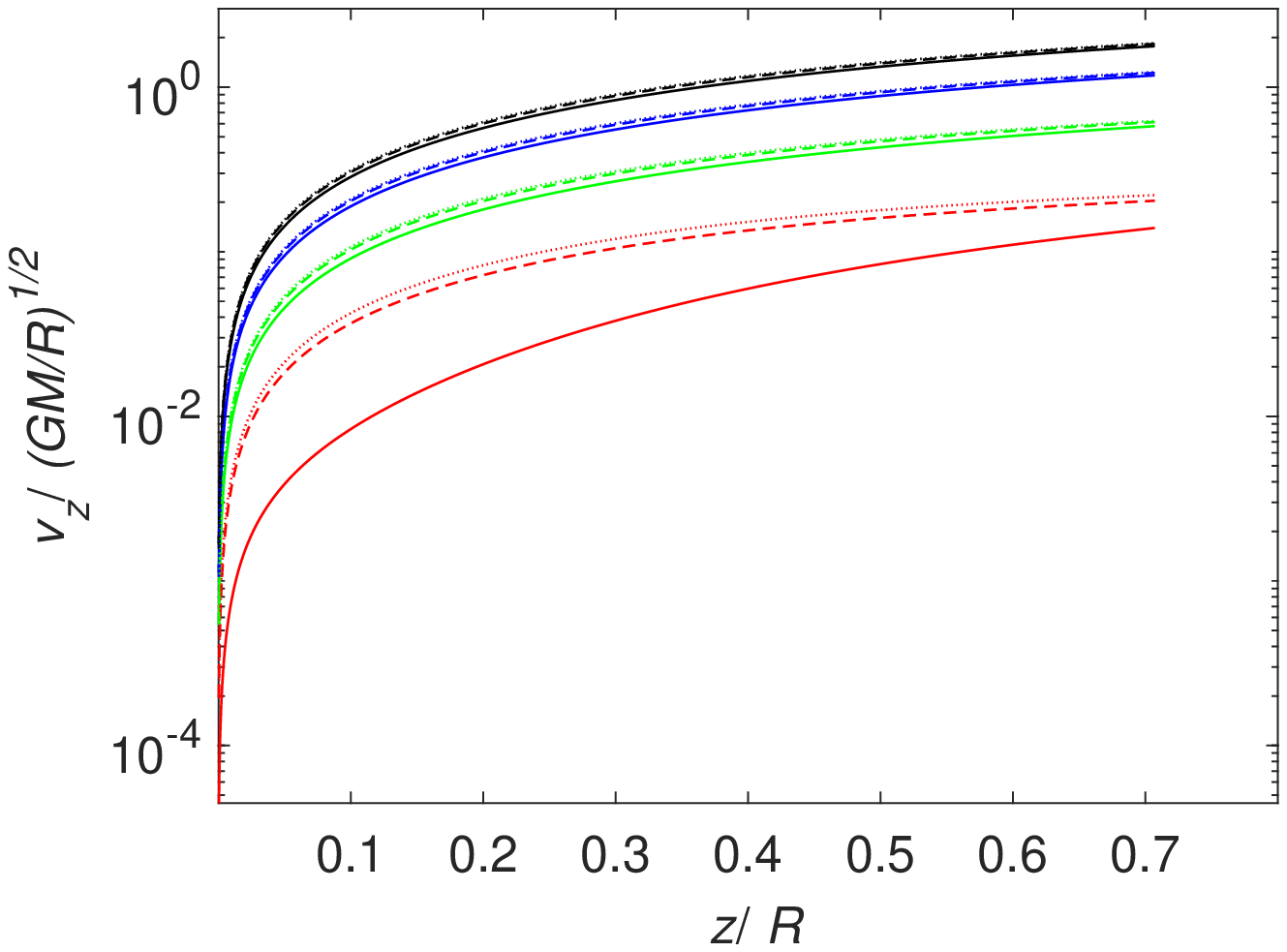}{0.4\textwidth}{}}
\caption{The dynamics of the gas motion in the vertical direction of the disk.
The left panel: the density distributions of the disk. The black lines indicate the results of $\tilde{f}_{\rm rad}=0.2\tilde{f}_{\rm rad}^{\rm crit}$ (solid), $0.5\tilde{f}_{\rm rad}^{\rm crit}$ (dashed) and $\tilde{f}_{\rm rad}^{\rm crit}$ (dotted) respectively. The colored lines are the results of the disk calculated with $\tilde{f}_{\rm rad}=2\tilde{f}_{\rm rad}^{\rm crit}$. The different values of the vertical optical depth, $\tau=100$ (red), $1000$ (green), and $5000$ (blue), are adopted respectively. The right panel: the vertical velocities of the gas in the disk calculated with different values of $\tau$ and $\tilde{f}_{\rm rad}$. The solid lines are the cases of $\tau=100$, while the dashed and dotted lines are for $\tau=1000$ and $5000$ respectively. The results with different values of $\tilde{f}_{\rm rad}$ are labeled with different colors, $\tilde{f}_{\rm rad}=1.1\tilde{f}_{\rm rad}^{\rm crit}$ (red), $2\tilde{f}_{\rm rad}^{\rm crit}$ (green), $5\tilde{f}_{\rm rad}^{\rm crit}$ (blue), and $10\tilde{f}_{\rm rad}^{\rm crit}$ (black), respectively.
\label{fig1}}
\end{figure}

We plot the vertical density distributions and the vertical motions of the gas in the disk with outflows in Figure \ref{fig1}. In the case of $\tilde{f}_{\rm rad}\le \tilde{f}_{\rm rad}^{\rm crit}$, the density distribution $\rho(z)$ is given by Equation (\ref{rho_z_nw}), while $\rho(z)$ is calculated from Equation (\ref{rho_tau}) together with the integration of Equation (\ref{tau_z}) in the disk region with outflows. With derived $\rho(z)$, the vertical motion of the gas in the disk is calculated with Equations (\ref{f_rad_z2}) and (\ref{vert_disc4}) (see the right panel in Figure \ref{fig1}). The gas is accelerated upwards when the radiation force is greater than the vertical component of the gravity, otherwise the gas is in vertical hydrostatic equilibrium. It is found that the vertical velocity of the gas at the disk surface can be as high as the Keplerian velocity provided $\tilde{f}_{\rm rad}\ga 5\tilde{f}_{\rm rad}^{\rm crit}$. {The results in Figure \ref{fig1} are calculated for given values of the dimensionless radiation flux $\tilde{f}_{\rm rad}$, which are independent of radius $R$.} {We plot the rates of the angular momentum carried away by the outflows in Figure \ref{fig2c}. }

\begin{figure}
	\centering
	\includegraphics[width=0.65\columnwidth]{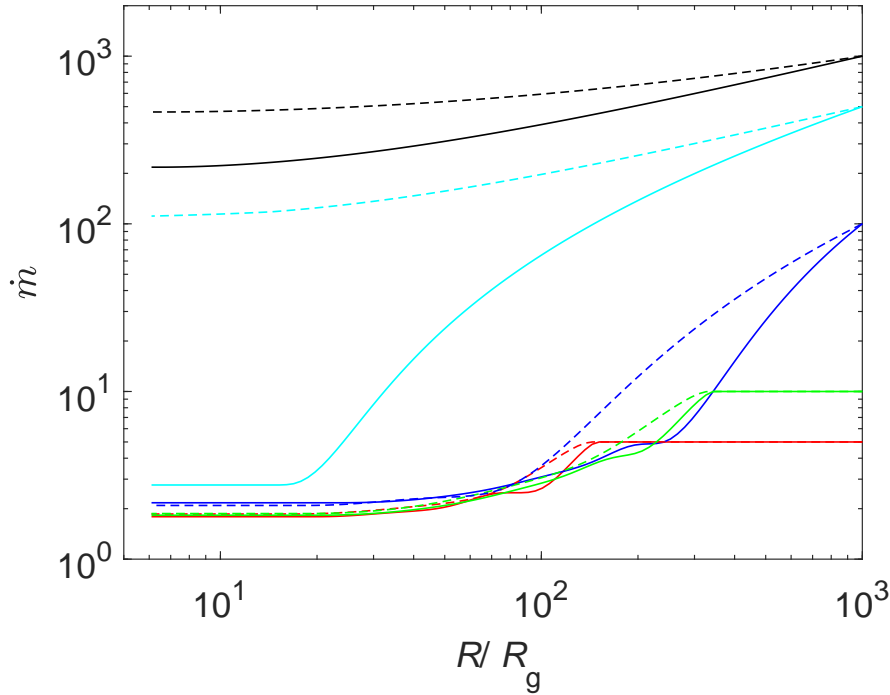}
	\caption{The accretion rates as functions of radius in the disks. The solid and dashed lines correspond to the results calculated with $\alpha=0.1$ and $0.3$ respectively. The colored lines indicate the results of $\dot{m}_{\rm out}=5$ (red), $10$ (green), $100$ (blue) and $500$ (cyan), and $1000$ (black), respectively. The outer disk radius $r_{\rm out}=1000$ is adopted. }
	\label{fig2}
\end{figure}

\begin{figure}
\gridline{\fig{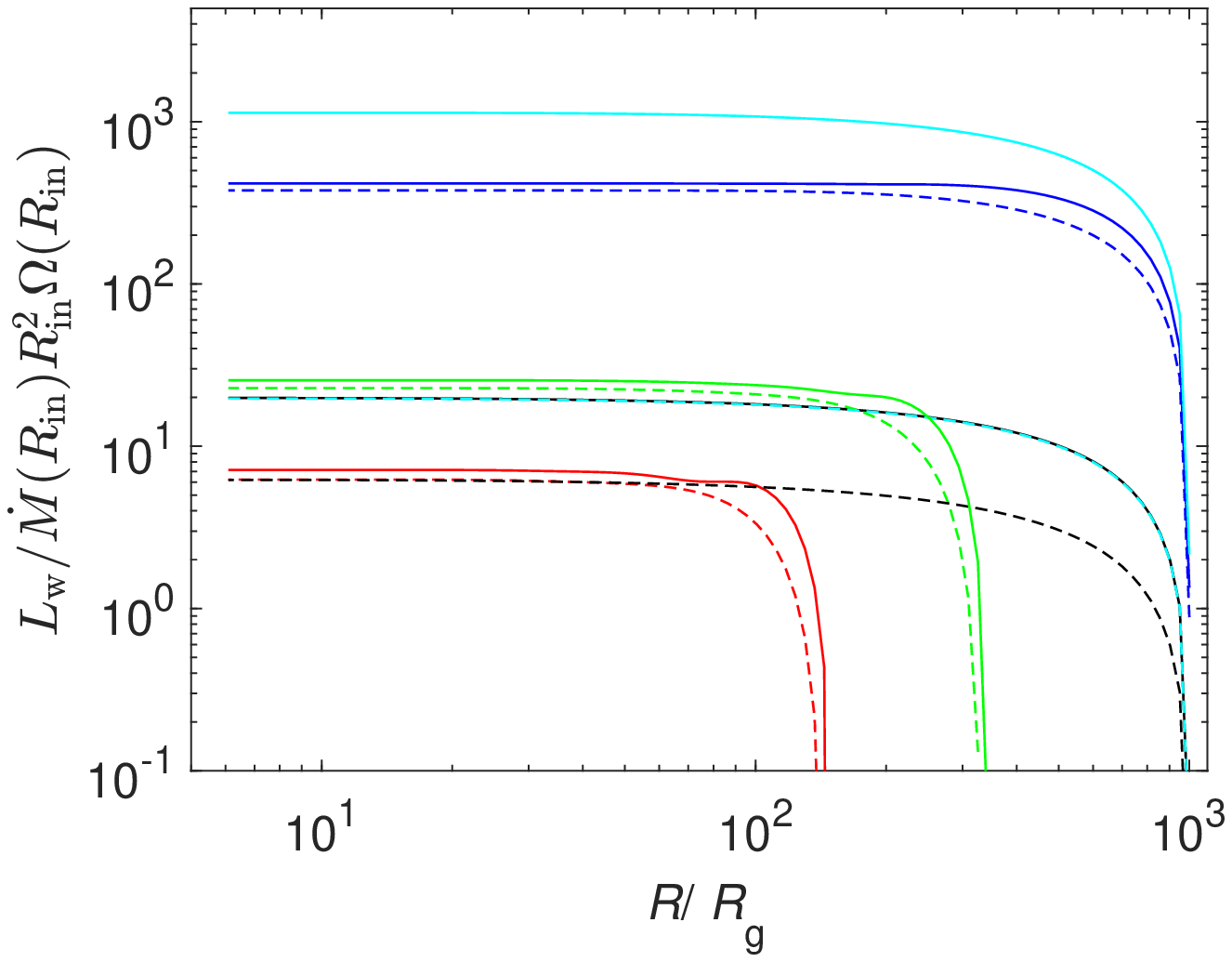}{0.4\textwidth}{}
 \fig{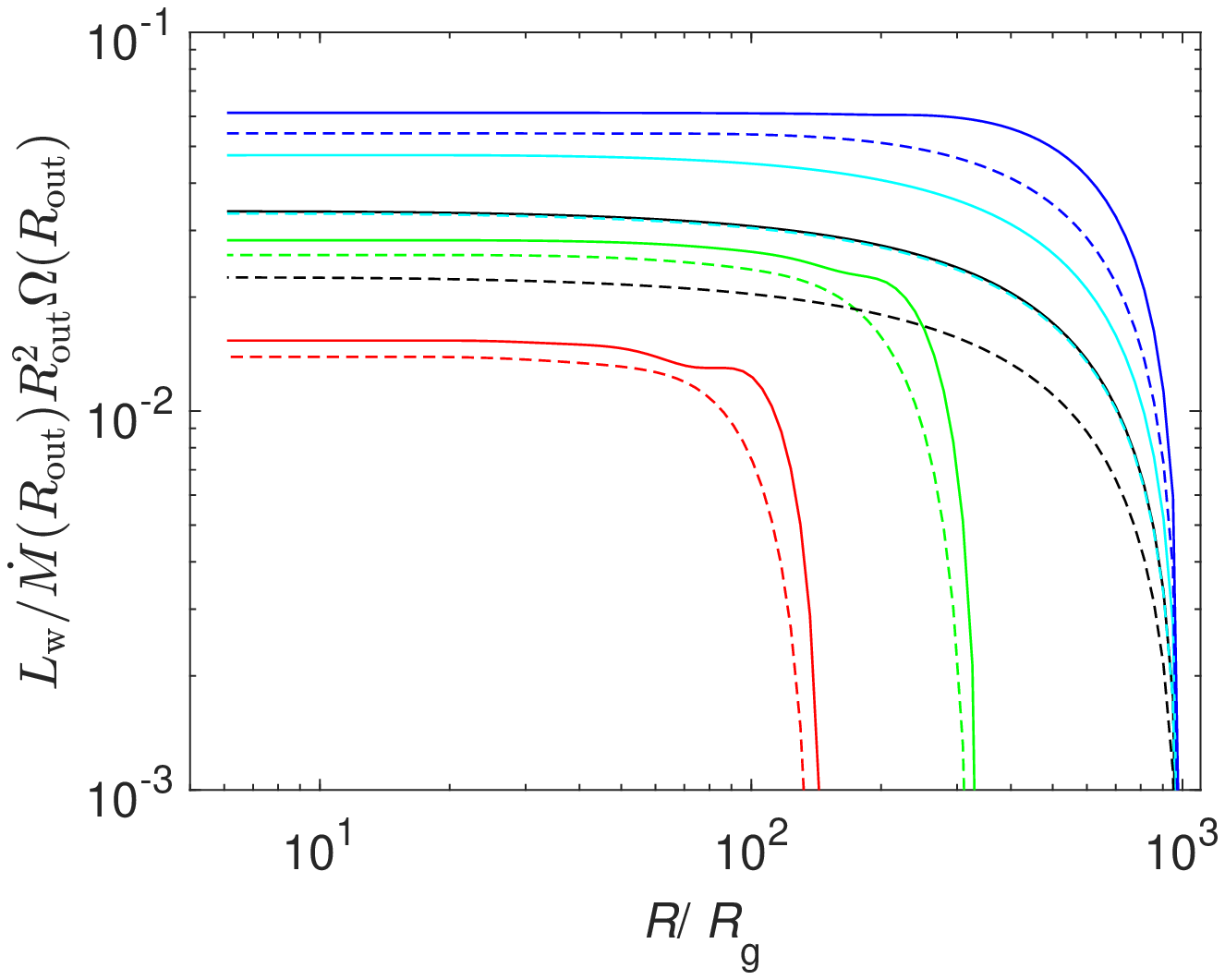}{0.4\textwidth}{}}
\caption{{The rates of the angular momentum carried away by the outflows from the disk region between $R$ and $R_{\rm out}$. The solid and dashed lines correspond to the cases with $\alpha=0.1$ and $0.3$ respectively. The colored lines are the results for different accretion rates, $\dot{m}(r_{\rm out})=5$ (red), 10 (green), 100 (blue), $500$ (cyan), and 1000 (black), respectively. The left panel contains the ratios of the angular momentum carried away by the outflows to that of the gas accreted by the BH, while the results normalized to the angular momentum rate of the gas accreted onto the outer edge of the disk $r_{\rm out}=1000$ are plotted in the right panel.
\label{fig2c}}}
\end{figure}

In Figure \ref{fig2}, we plot the accretion rate varying with radius for different values of the accretion rate at the outer radius of the disk. Most gas inflowing onto the outer edge of the disk is driven into outflows, and only a small fraction of gas is finally swallowed by the BH, if the accretion rate $\dot{m}(r_{\rm out})\la 100$. In the case of an extremely high accretion rate, i.e., $\dot{m}(r_{\rm out})\sim 1000$, nearly half of the gas is finally accreted by the BH. It is found that the outflow properties (e.g., the mass loss rate in the outflows) vary with the value of the viscosity parameter $\alpha$.

\begin{figure}
	\centering
	\includegraphics[width=0.65\columnwidth]{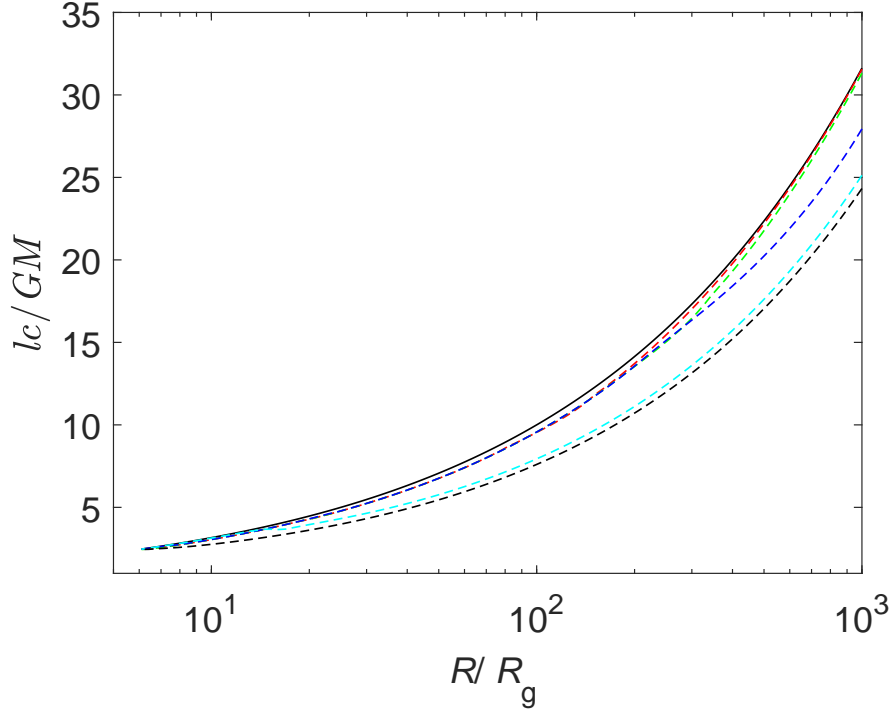}
	\caption{The specific angular momenta of the gas in the disks as functions of radius. The solid line indicates the angular momentum of the Keplerian rotation, while the dashed lines are the specific angular momentum of the disk with different values of model parameters. The colored lines indicate the results of $\dot{m}_{\rm out}=5$ (red), $10$ (green), $100$ (blue) and $500$ (cyan), and $1000$ (black), respectively. The outer disk radius $r_{\rm out}=1000$, and the viscosity parameter $\alpha=0.1$, are adopted in the calculations.}
	\label{fig2b}
\end{figure}

The angular velocity of a supercritical accretion disk may deviate from the Keplerian  value. We plot the specific angular momentum of the disk varying with radius in Figure \ref{fig2b}. It is found that the disk rotation deviates substantially from the Keplerian rotation if the accretion rate $\dot{m}(r_{\rm out})\ga 100$ at the outer radius, while it is very close to the Keplerian value in the case of $\dot{m}(r_{\rm out})\la 10$.

\begin{figure}
\gridline{\fig{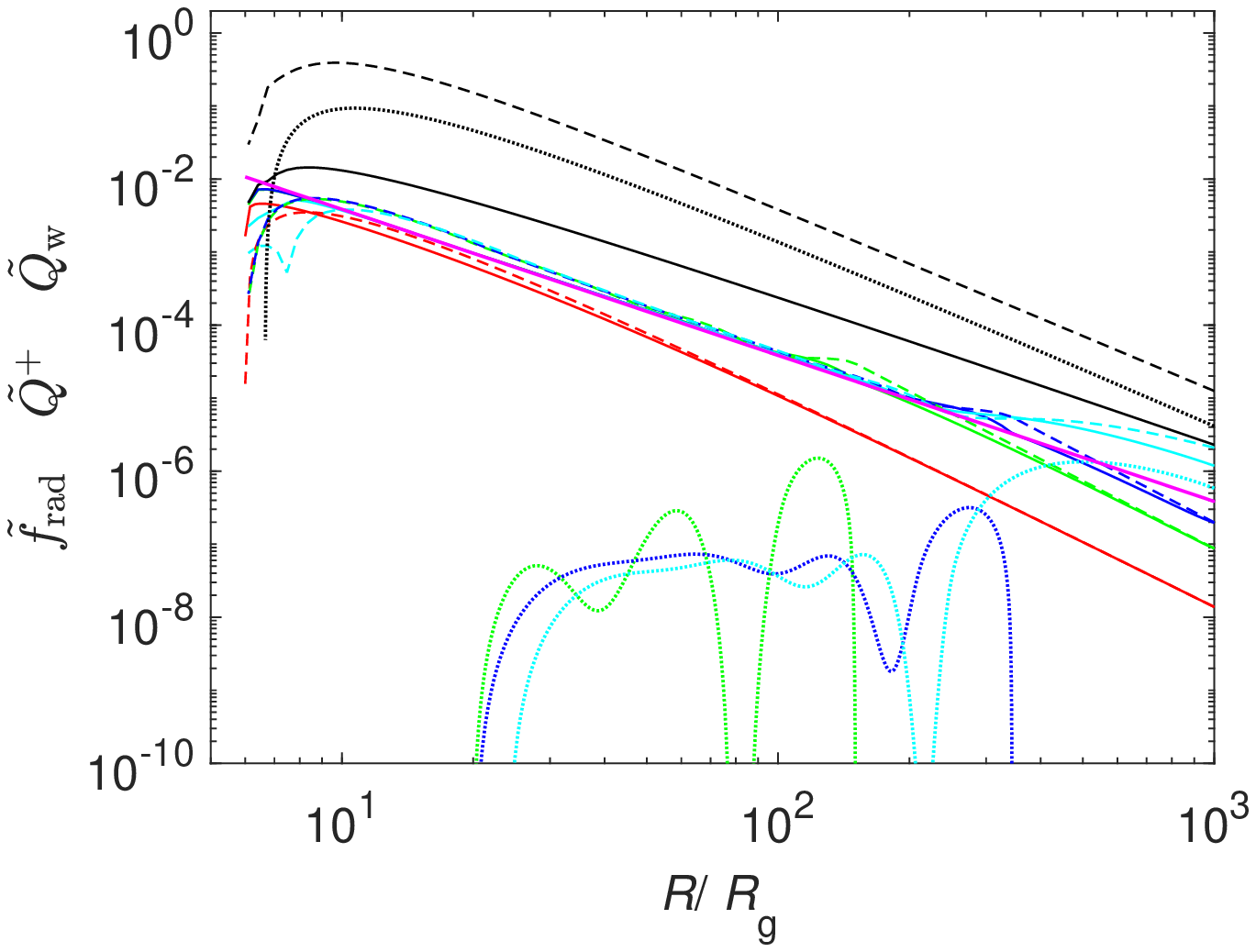}{0.4\textwidth}{}
 \fig{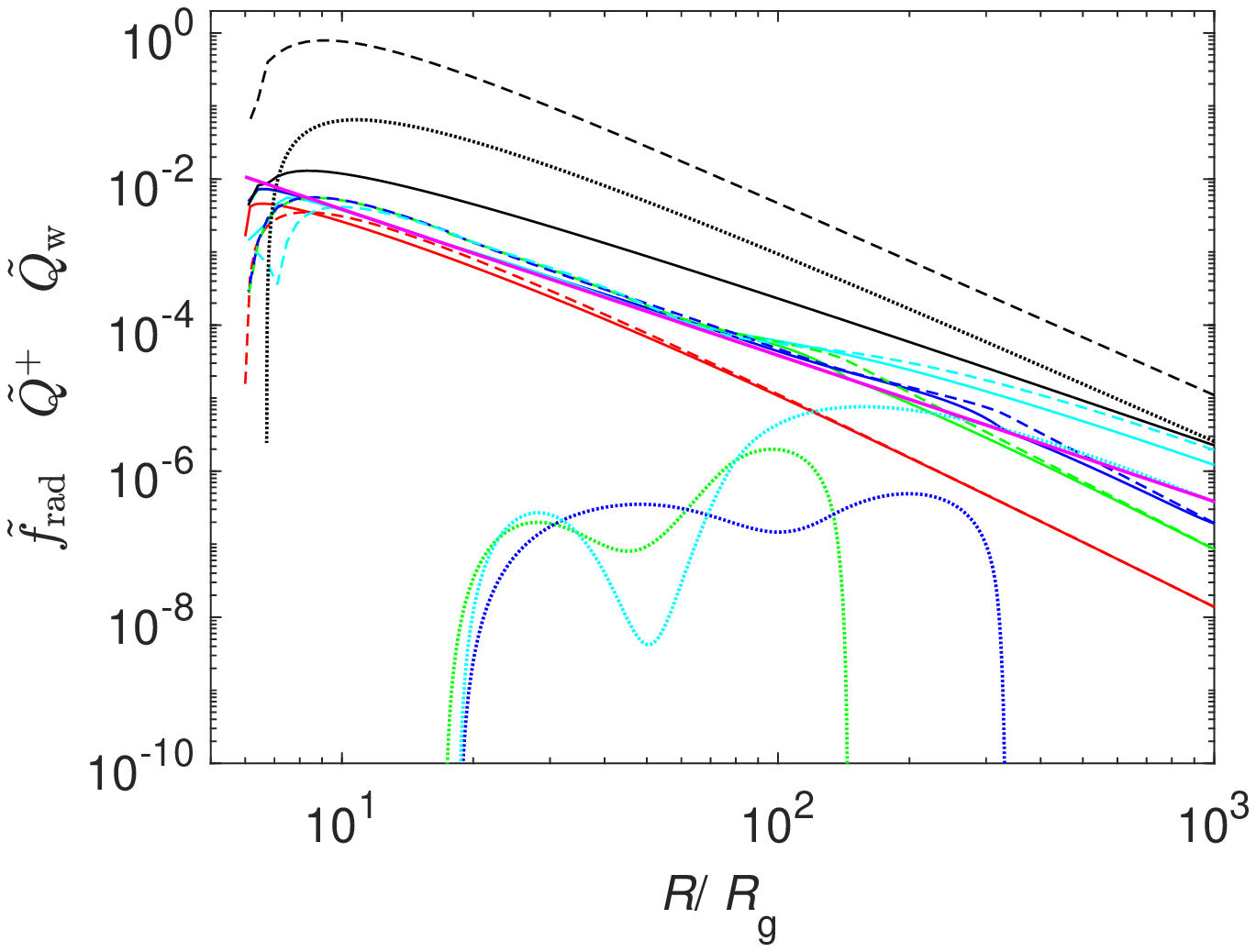}{0.4\textwidth}{}}
\caption{The viscously dissipated power, radiation flux and the kinetic power tapped into the outflows in the disk as functions of radius. The results calculated with $\alpha=0.1$ are plotted in the left panel, while the right panel is for the results with $\alpha=0.3$. The colored lines are the results for different accretion rates, $\dot{m}(r_{\rm out})=1$ (red), 5 (green), 10 (blue), 100 (cyan) and 1000 (black), respectively. The radiation flux $\tilde{f}_{\rm rad}$ (solid lines), the viscously dissipated power  $\tilde{Q}^+$ (dashed lines), and the kinetic power of the outflows $\tilde{Q}_{\rm w}$ (dotted lines) from the unit area of the disk vary with radius with different values of the parameters are plotted. The magenta lines correspond to $\tilde{f}_{\rm rad}=\tilde{f}_{\rm rad}^{\rm crit}\equiv 2\sqrt{3}/9r^2$. {We note red dotted line is absent because $\tilde{f}_{\rm rad}<\tilde{f}_{\rm rad}^{\rm crit}$ is always satisfied in the disk, i.e., no outflow is driven from the disk, when $\dot{m}(r_{\rm out})=1$.}
\label{fig3}}
\end{figure}

\begin{figure}
	\centering
	\includegraphics[width=0.65\columnwidth]{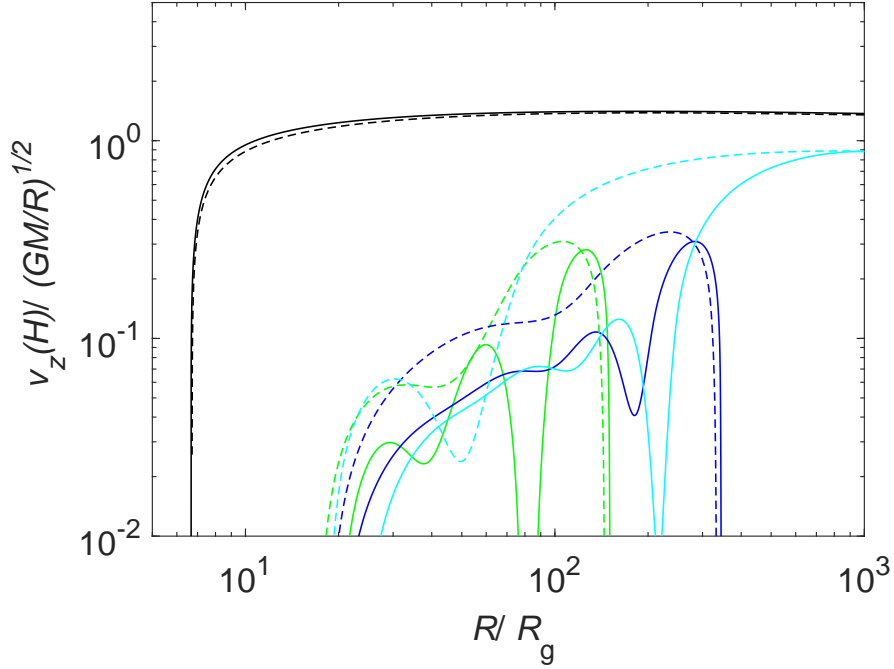}
	\caption{The vertical velocity $v_z(H)$ of the gas at the disk surface $z=H$. {The results calculated with $\alpha=0.1$ are plotted as solid lines, while the dashed lines are for the results with $\alpha=0.3$. The colored lines are the results for different accretion rates, $\dot{m}(r_{\rm out})=5$ (green), 10 (blue), 100 (cyan) and 1000 (black), respectively.} }
	\label{fig3b}
\end{figure}

The viscously dissipated power, the radiation flux, and the power tapped into the outflows from the unit area of the disk surface are plotted in Figure \ref{fig3}. The kinetic power of the vertical motion of the gas is usually much less than the radiation flux, except the cases with extremely high accretion rates $\dot{m}(r_{\rm out})\sim 10^3$. It implies that the structure of the disk is affected by outflows predominantly through their impact on the accretion rate, i.e., the accretion rate in the disk is mainly regulated by the outflows, provided the accretion rate is not very high at the outer edge of the disk. In Figure \ref{fig3b}, we plot the gas velocity of the vertical motion at the disk surface. It is found that the velocity $v_z(H)\ga (GM/R)^{1/2}$ only for the disk accreting at an extremely high rate, which implies the gas leaving from the disk should be further accelerated by the radiation force of the disk in order to escape to infinity.

\begin{figure}
	\centering
	\includegraphics[width=0.65\columnwidth]{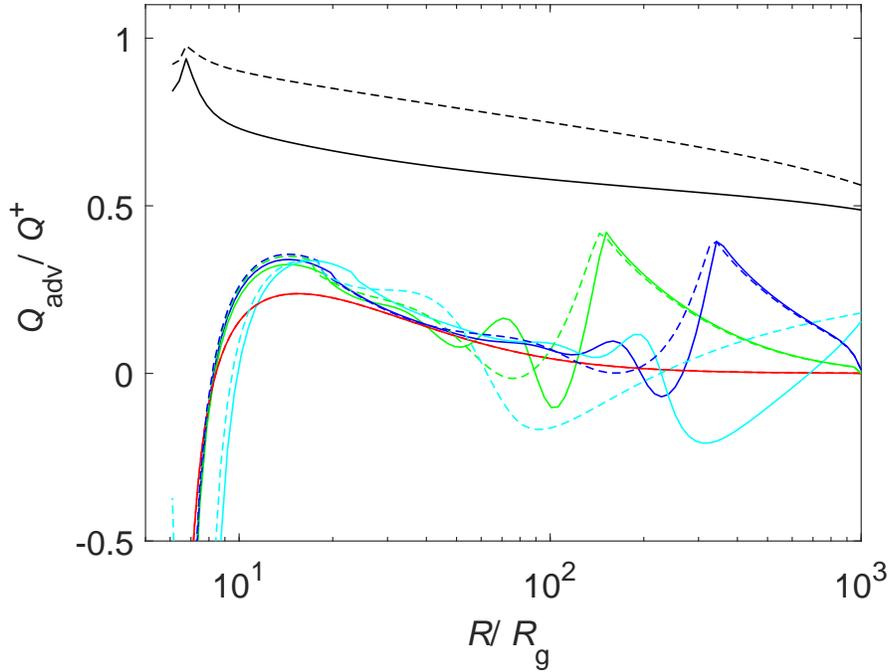}
	\caption{The fraction of radial energy advection $f_{\rm adv}=Q_{\rm adv}/Q^+$ varying with the radius of the disk. {The results calculated with $\alpha=0.1$ are plotted as solid lines, while the dashed lines are for the results with $\alpha=0.3$. The colored lines are the results for different accretion rates, $\dot{m}(r_{\rm out})=1$ (red), 5 (green), 10 (blue), 100 (cyan) and 1000 (black), respectively.} }
	\label{fig4}
\end{figure}

\begin{figure}
	\centering
	\includegraphics[width=0.65\columnwidth]{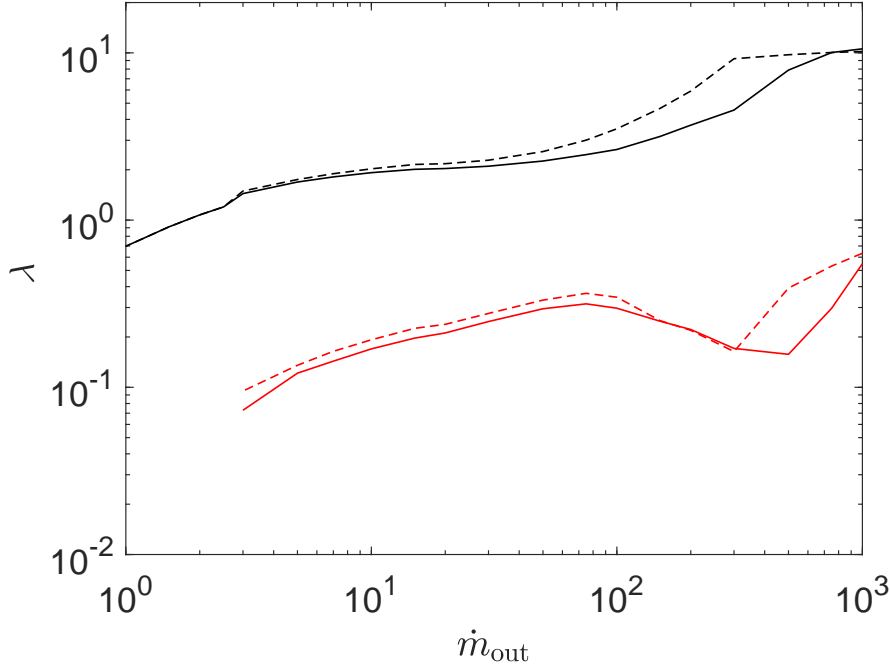}
	\caption{The Eddington ratios of the disk as functions of the accretion rate at the outer radius of the disk $r_{\rm out}=1000$ (black lines). The red lines indicate the least radiation power required to accelerate the gas from the disk surface to infinity. The solid lines indicate the results calculated with $\alpha=0.1$, while the dashed lines are for $\alpha=0.3$. }
	\label{fig5}
\end{figure}

\begin{figure}
\gridline{\fig{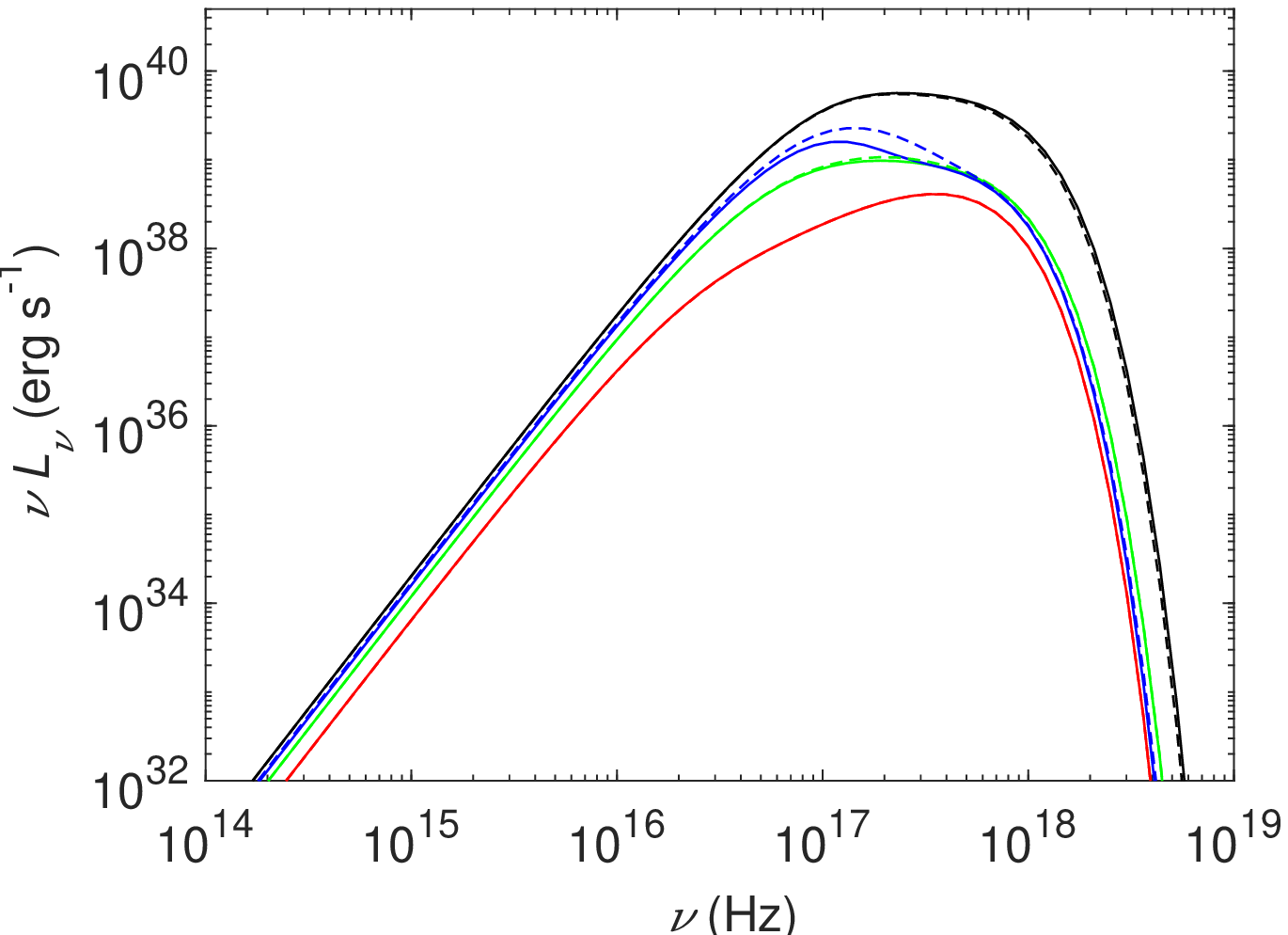}{0.4\textwidth}{}
 \fig{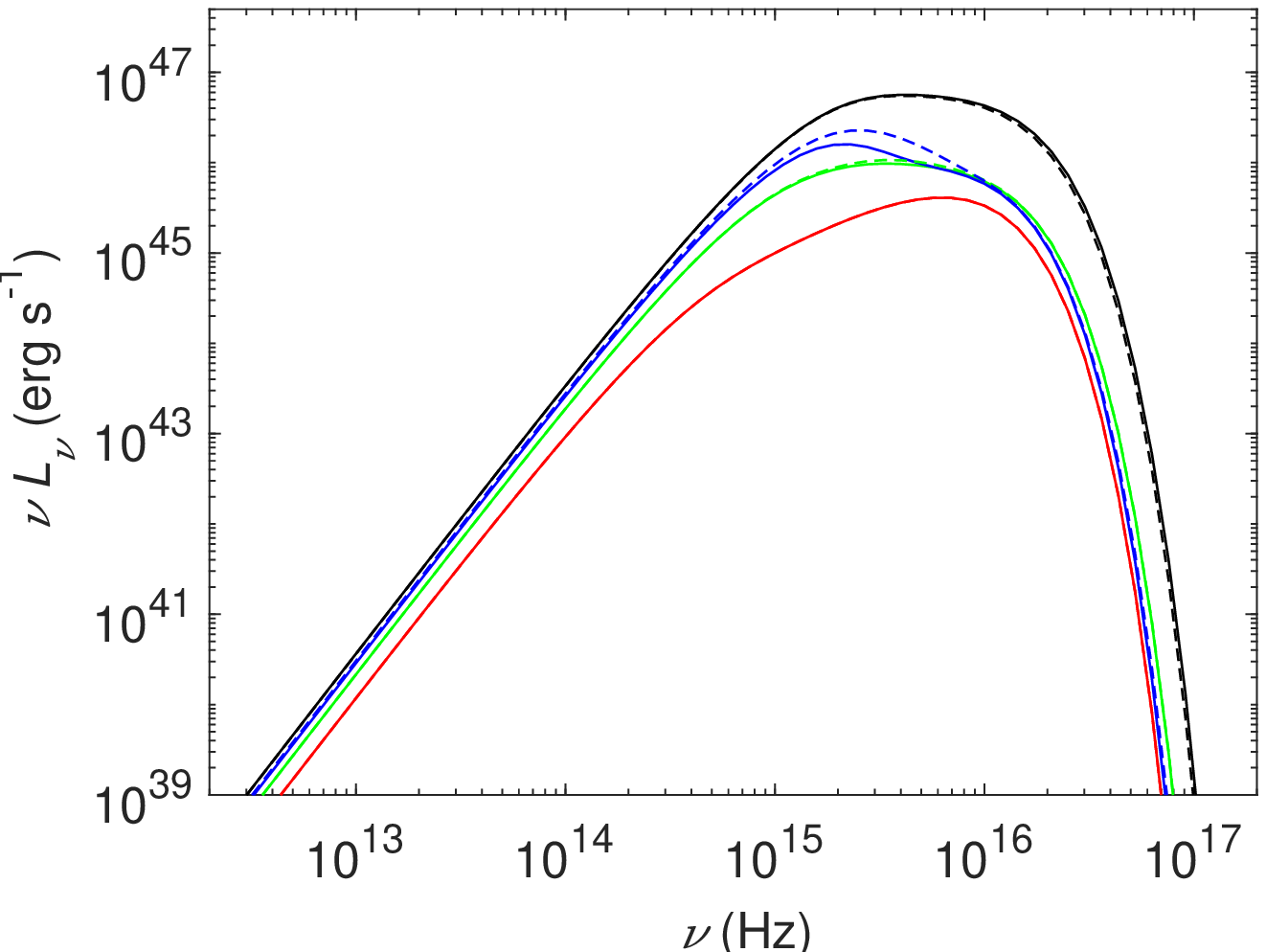}{0.4\textwidth}{}}
\caption{The continuum spectra of the disks (left panel: $M=10 M_\odot$, right panel: $M=10^8 M_\odot$). The solid lines are the results calculated with $\alpha=0.1$, while the dashed lines are for $\alpha=0.3$. The colored lines correspond to different accretion rates at the outer radius $r_{\rm out}=1000$: $\dot{m}(r_{\rm out})=1$ (red), 10 (green), 100 (blue), and 1000 (black), respectively.
\label{fig6}}
\end{figure}

The fraction of the radial energy advection in the disk is plotted in Figure \ref{fig4}. It is found that the fractions of the advection $f_{\rm adv}=Q_{\rm adv}/Q^+$ are large in the extremely high accretion rate cases. We find a non-linear relation of the Eddington-scaled luminosity of the disk with the accretion rate (see Figure \ref{fig5}). The outflows are present only when the accretion rate is sufficiently high, while the outflow is completely suppressed in the whole disk if the accretion rate is low $\dot{m}(r_{\rm out})\la 3$. {For the gas with vertical velocity at the disk surface $v_z(H)$ being substantially lower than $\sim(GM/R)^{1/2}$, the extra-power is needed to accelerate the gas in the outflows to infinity.} We estimate the least power of the disk radiation required to accelerate the gas from the disk surface to infinity with particle approximation in Figure \ref{fig5}, {i.e., the gas is assumed as a particle moving to infinity}. It is found that the least power required is usually at an order of one magnitude lower of the radiation power of the disk. The spectra of the disks surrounding a stellar mass BH ($M=10M_\odot$) or a super-massive BH with $10^8 M_\odot$ are plotted in Figure \ref{fig6} for different values of accretion rate.

\begin{figure}
	\centering
	\includegraphics[width=0.65\columnwidth]{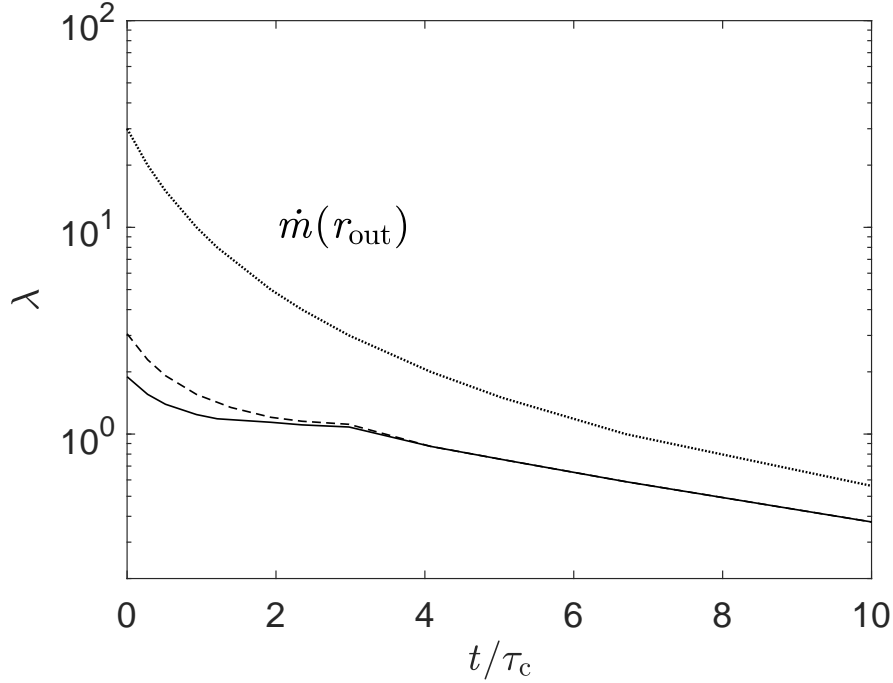}
	\caption{The accretion rate $\dot{m}_{\rm out}$ and the Eddington ratio $\lambda$ evolve with time in the TDEs. The dotted line indicates the accretion rate at the outer radius of the disk evolving with time as $\propto t^{-5/3}$. The solid line is the light curve of the disk emission with $\alpha=0.1$, while the dashed line is for the disk with $\alpha=0.3$. The outer radius of the disk $r_{\rm out}=100$ is adopted in the calculations. }
	\label{fig7}
\end{figure}

\begin{figure}
	\centering
	\includegraphics[width=0.65\columnwidth]{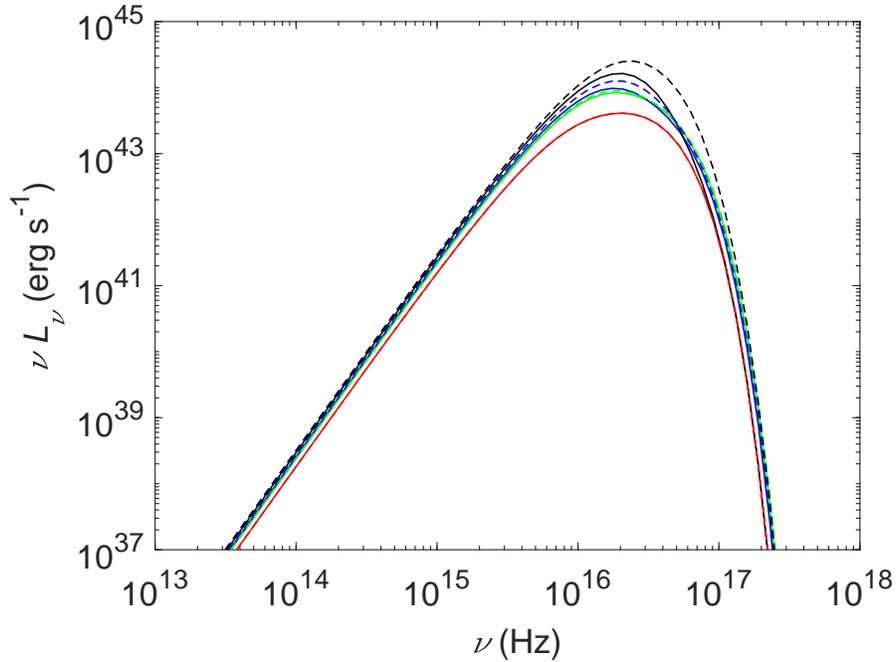}
	\caption{The continuum spectra of the disks surrounding BHs with $M=10^6 M_\odot$. The solid lines indicate the disk spectra calculated with $\alpha=0.1$, while the dashed lines are for the spectra calculated with $\alpha=0.3$. The colored lines correspond to different accretion rates at the outer radius $r_{\rm out}=100$: $\dot{m}(r_{\rm out})=1$ (red), 5 (green), 10 (blue), and  30 (black), respectively.}
	\label{fig8}
\end{figure}

We apply our model to the observations of TDEs. After a star is disrupted by the tidal force of the BH with $\sim 10^6M_\odot$, a disk with size of several hundred gravitational radii is formed and its accretion rate is predicted to decline with time as $\dot{M}\propto t^{-5/3}$ \citep*[][]{1988Natur.333..523R}. We assume that the accretion rate of disk fed by the TDE follows
\begin{equation}
\dot{m}_{\rm out}(t)=\left({\frac {t+\tau_{\rm c}}{\tau_{\rm c}}}\right)^{-5/3}\dot{m}_{\rm out}(t=0),\label{mdot_t}
\end{equation}
{where $\tau_{\rm c}$ is the characteristic timescale of the accretion rate decay in the TDE.} The light curve of such an accretion disk with outflows is then calculated with our model
(see Figure \ref{fig7}). In the calculations of the light curves, we adopt the initial accretion rate $\dot{m}_{\rm out}(t=0)=30$ at $r_{\rm out}=100$. The light curve derived in this work is qualitatively consistent with the slowly declining X-ray light curves of the TDEs \citep*[see e.g., Figure 5 in][]{2016MNRAS.455.2918H}. The disk spectra of a TDE with different values of accretion rate are plotted in Figure \ref{fig8}.

\begin{figure}
\gridline{\fig{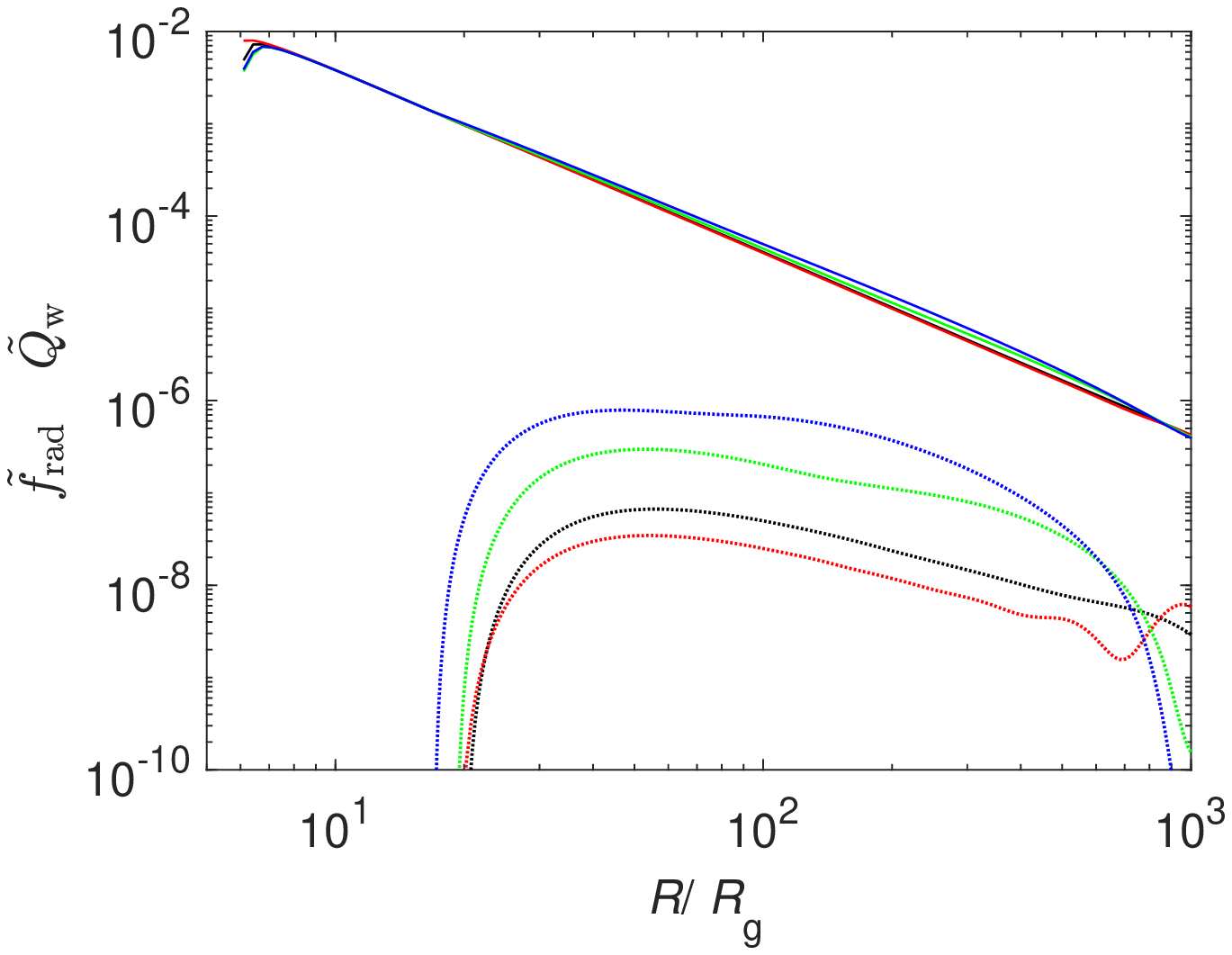}{0.4\textwidth}{}
 \fig{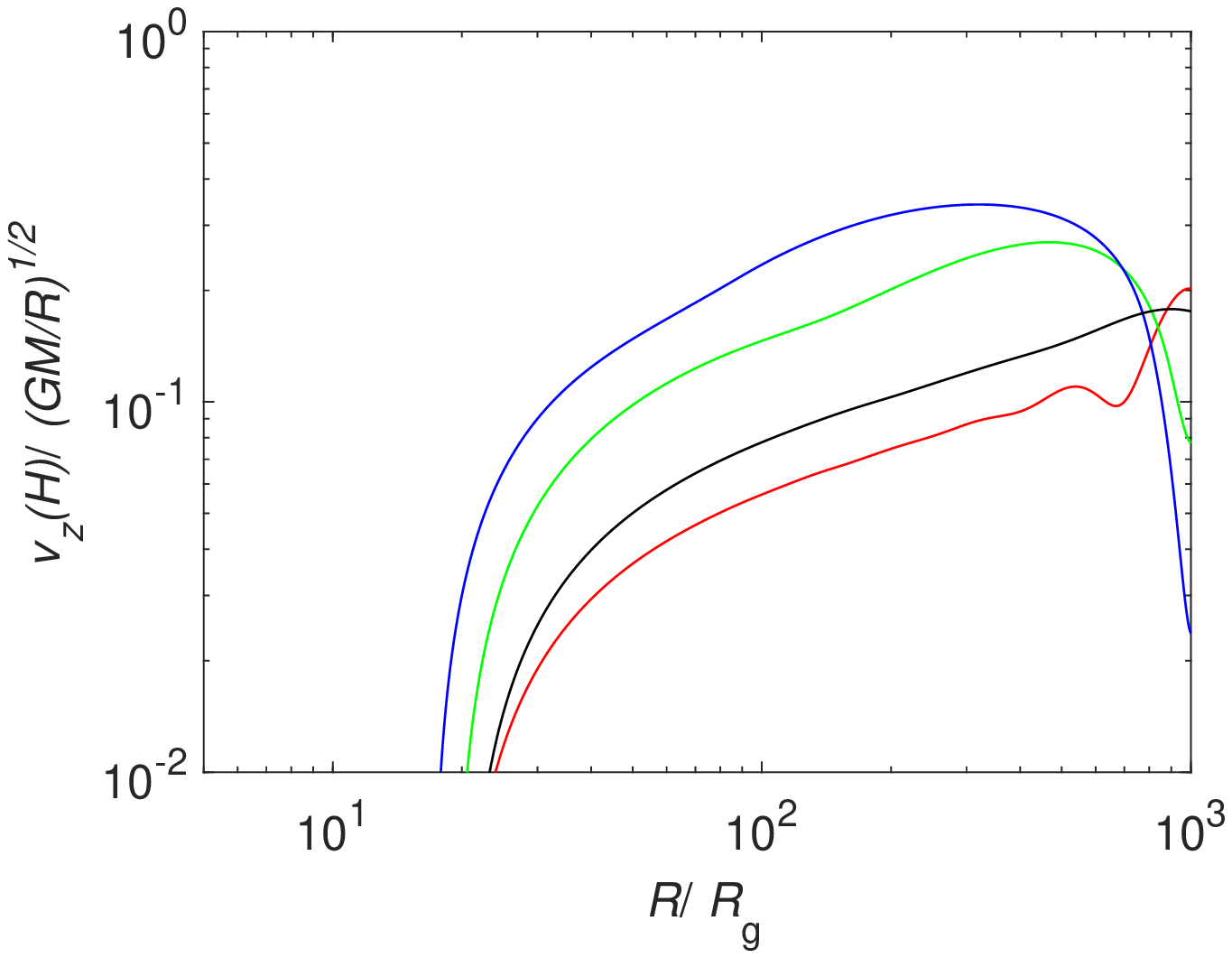}{0.4\textwidth}{}}
\caption{{The properties of the disk-outflow systems with different values of $\alpha$. The black lines are the results with $\alpha=0.1$, while the colored lines indicate the results for  $\alpha=0.05$ (red), $0.3$ (green) and $0.5$ (blue). The calculations are carried out with $\dot{m}(r_{\rm out})=20$ and $r_{\rm out}=1000$. The left panel: the radiation flux $\tilde{f}_{\rm rad}$ (solid lines) and the kinetic power $\tilde{Q}_{\rm w}$ tapped into the outflows (dotted lines) in the disk as functions of radius. The right panel: the vertical velocity $v_z(H)$ of the gas at the disk surface $z=H$.}}
\label{fig9}
\end{figure}

\begin{figure}
\gridline{\fig{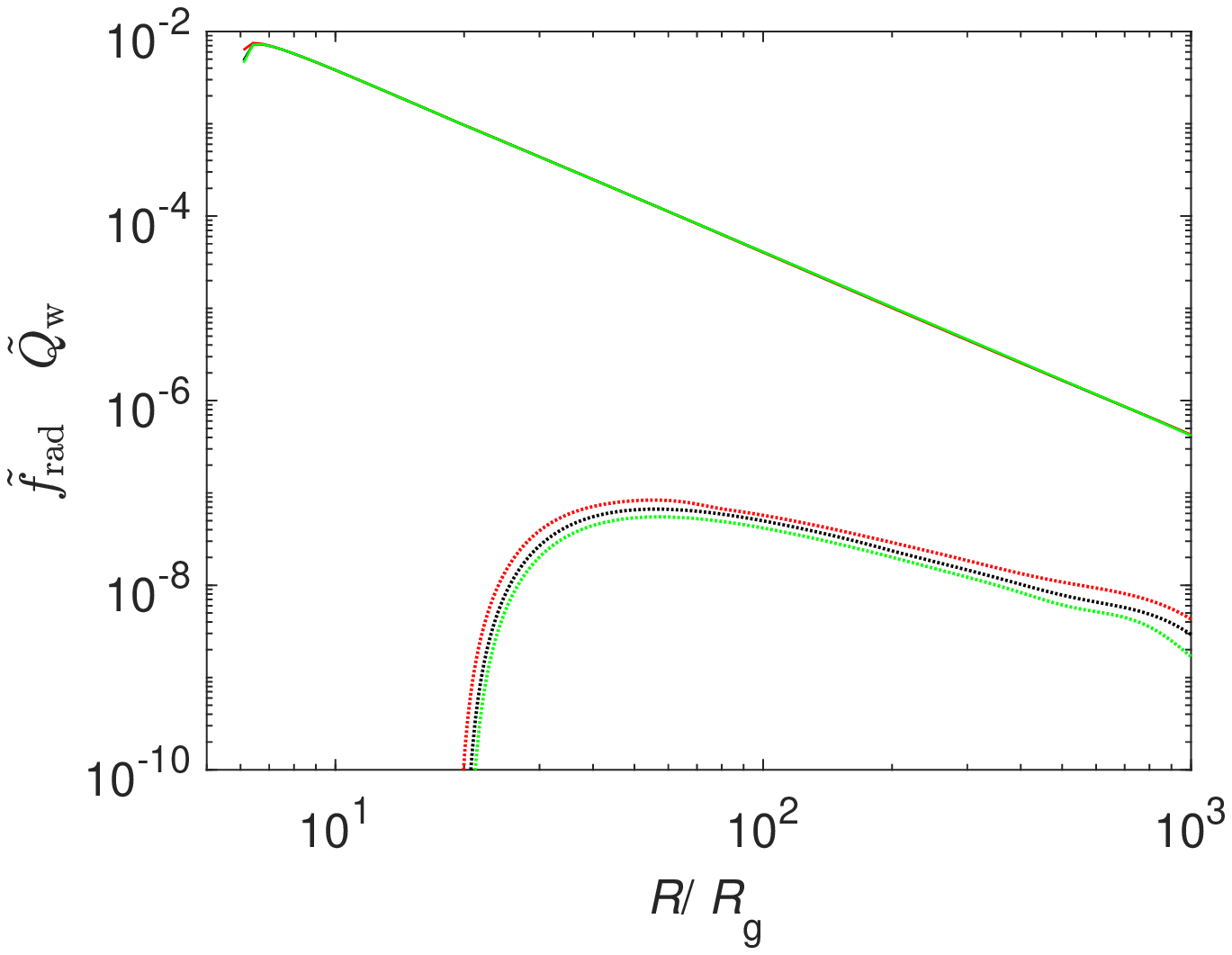}{0.4\textwidth}{}
 \fig{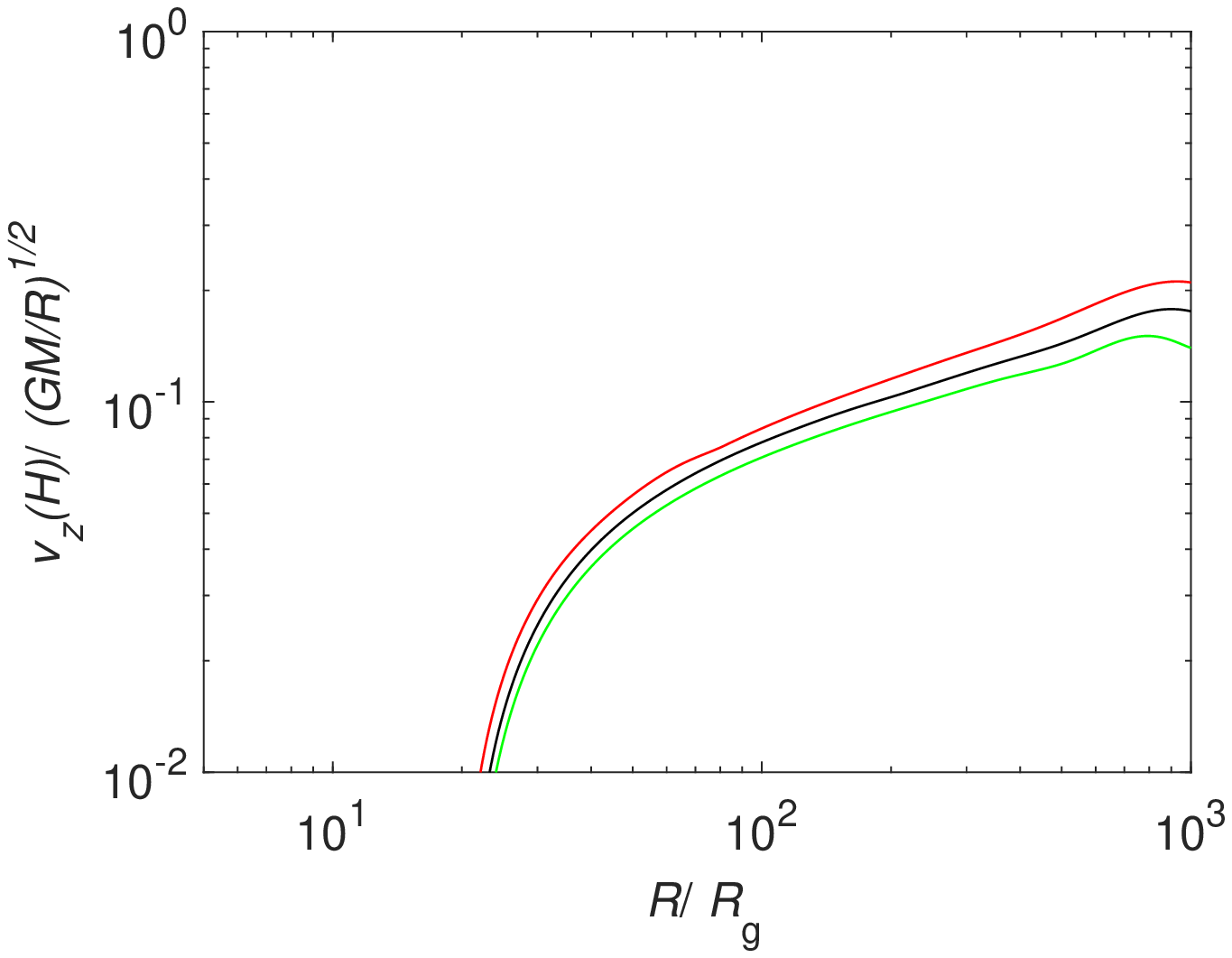}{0.4\textwidth}{}}
\caption{{The properties of the disk-outflow systems with different values of $\gamma$. The black lines are the results with $\gamma=1.5$, while the red and green lines are for $\gamma=1.4$ and $1.6$ respectively. The calculations are carried out with $\dot{m}(r_{\rm out})=20$, $r_{\rm out}=1000$, and $\alpha=0.1$. The left panel: the radiation flux $\tilde{f}_{\rm rad}$ (solid lines) and the kinetic power $\tilde{Q}_{\rm w}$ tapped into the outflows (dotted lines) in the disk as functions of radius. The right panel: the vertical velocity $v_z(H)$ of the gas at the disk surface $z=H$.}}
\label{fig10}
\end{figure}

{We plot the properties of the disk-outflow systems with different values of the viscosity parameter $\alpha$ in Figure \ref{fig9}, while the impacts of the value of polytropic index $\gamma$ are given in Figure \ref{fig10}. }


\vskip 1cm

\section{Discussion}\label{discussion}

The gas in the accretion disk is in vertical hydrostatic equilibrium when the radiation flux is lower than the critical value. The density distribution $\rho(z)$ is nearly constant in $z$-direction, i.e., vertically homogeneous, with a sharp decrease of density at $z\sim H$, provided $z/R\ll 1$, which is consistent with the standard thin accretion disk in the previous works \citep*[][]{1973A&A....24..337S,2004ApJ...605L..45T}. The density distribution deviates from a homogeneous distribution when the disk thickness increases. It declines along the vertical direction with a sharp decrease at $z\sim H$. When the outgoing radiation flux approaches a critical value, the gas density ceases to zero at $z=H=\sqrt{2}R/2$ (see Figure \ref{fig1}). Increasing the radiation flux further above the critical value leads to the vertical radiation force always being higher than the vertical component of the BH gravity, i.e., vertical equilibrium between the radiation force and gravity can no longer be maintained. In this case, the gas in the disk would inevitably move upwards to the disk surface to form outflows. The mass loss rate in the outflows is the product of the density and velocity of the gas at the disk surface, which are available by solving the dynamical equation of vertical gas motion. The velocity of the gas at the disk surface increases with the radiation flux, which means that both the velocity and mass loss rate in the outflows increases with the radiation flux (Figure \ref{fig1}).

Outflows will be inevitably driven from the disk when the accretion rate is high, while no outflow appears in the low accretion rate case. It is obvious that the mass loss rate would increase with radiation flux, which is governed by the accretion rate, though non-linearly, because of the radial energy advection/kientic power of gas tapped into outflows. We find that the accretion rate at the inner edge of the disk (i.e., the rate of the mass accreted by the BH) remains roughly around $2-3 \dot{M}_{\rm Edd}$ if the accretion rate at the outer radius of the disk is up to a moderately high rate with $\sim 100\dot{M}_{\rm Edd}$, which is consistent with the previous works \citep*[][]{2015MNRAS.448.3514C,2019ApJ...885...93F}. In the slim disk, no outflow is considered, and its radiation is mainly governed by the accretion rate and the radial energy advection \citep*[][]{1988ApJ...332..646A}. In this work, the radiation driven outflows may carry away a substantial fraction of the gas fed at the outer radius of the disk, which means the disk structure, and then the radiation of the disk, are significantly altered by the outflows. We find that the term of power tapped into the outflows is always negligible except in the case of the  extremely high accretion rate with hundreds of Eddington rate (see Figure \ref{fig3}). The radial energy advection has been properly included in our calculations, however, we find that the fraction of advection is not very high, which implies the disk structure is predominantly determined by the accretion rate regulated by the outflows. The disk structure is predominantly impacted by the outflows by regulating the accretion rate in the disk, which is quite different from the normal slim accretion disk \citep*[][]{1988ApJ...332..646A}. Most of the gas feeding the accretion disk at its outer radius is removed by the outflows in the cases with accretion rates up to moderately high values, which is qualitatively consistent with the numerical simulations \citep*[][]{2011ApJ...736....2O,2014ApJ...780...79Y,2014ApJ...796..106J,2018ApJ...867..100Y,2019ApJ...880...67J,2020MNRAS.497..302T}.
{Accordingly, the rates of the angualr mementum carried away by the outflows are usually much higher than that of the gas accreted by the BH, while they are only a small fraction of that of the gas accreted onto the outer edge of the disk (see Figure \ref{fig2c}).}

The properties of the outflows are sensitive to the disk radiation, while the disk structure is mainly regulated by the mass loss in the outflows, and therefore the final rate of gas swallowed by the BH is determined by the coupled mechanism between the accretion disk and outflows. In the case of an extremely high accretion rate (several hundred $\dot{M}_{\rm Edd}$ or even higher), the outflows are very strong, $Q_{\rm w}$ becomes dominant over the radiation and advection terms in the energy equation (\ref{energy}) (Figure \ref{fig3}), which plays an important role on reducing the disk luminosity \citep*[][]{2016ApJ...821..104Y}. As the outflows are driven by the disk radiation, the rate of mass loss in the outflows in this case is therefore limited, i.e., the outflows are unable to remove a substantial fraction of the gas from the disk, and one finds that the rate of the mass accreted by the BH can still be very high, compared with the cases of the moderately high accretion rates with $\la 100\dot{M}_{\rm Edd}$. {When the accretion rate is low, the outflows are usually  driven from an annular region in the disk with an outer radius much less than the disk size. The width of this ring decreases with decreasing $\dot{m}(r_{\rm out})$, and it disappears when $\dot{m}(r_{\rm out})\la 2.5$. The disk structure in the ring is predominantly determined by the radial energy advection and the accretion rate regulated by the outflows, while only the advection plays an important role in the region outside the ring. {It is interesting to note that the outflows, in some cases with a low $\alpha$, are driven from two separated regions in the disk when the disk is accreting at a low or moderate rate (see the left panel in Figure \ref{fig3}). In the low accretion rate case, the radiation flux $f_{\rm rad}$ in the outer outflow driving disk region is only slightly higher than $f_{\rm rad}^{\rm crit}$, and the accretion rate decreases inwards due to the outflows, which leads to $f_{\rm rad}<f_{\rm rad}^{\rm crit}$ and the outflows being suppressed. The gas in the disk moves inwards further without outflows, then its radiation flux becomes $f_{\rm rad}>f_{\rm rad}^{\rm crit}$ again, and the outflows are driven from the inner region of the disk. In the case of $\dot{m}(r_{\rm out})=100$, the radiation flux $f_{\rm rad}$ is significantly higher than $f_{\rm rad}^{\rm crit}$ in the outer disk region, which drives a substantial fraction of the gas into the outflows. The radiation flux is therefore reduced to a value lower than $f_{\rm rad}^{\rm crit}$ due to a sharp accretion rate decline (see Figure \ref{fig2}), and the outflows are suppressed. Similar to the low accretion rate case, the outflows reappear where the radiation flux increases above $f_{\rm rad}^{\rm crit}$ again at an inner radius.}


{The rotation of the gas in a slim disk is sub-Keplerian \citep*[][]{1988ApJ...332..646A}. The situation is similar for the supercritical accretion disk considered in this work (see Figure \ref{fig2b}). The rotation of the gas in the disk deviates significantly from the Keplerian rotation when the accretion rate is high, while its angular velocity is very close to the Keplerian value if $\dot{m}\la 10$. We find that the specific angular momentum of the disk approaches the Keplerian value at the inner disk edge, which is caused by the zero-torque assumption at the inner disk edge (see Equation \ref{v_phi3}). The inner region of a supercritical accretion disk can be properly described by a global solution to the disk with outflows, which is beyond the scope of this work. }

The gas leaving the disk can be further accelerated by the radiation of the whole disk (not limited to the local radiation where the gas leaves the disk surface), The dynamics of the outflows can be calculated in principle when the radiation of the disk is known, which is a global problem, and is beyond the scope of this paper. It is evident that the gas with velocity $\sim (GM/R)^{1/2}$ at the disk surface is able to overcome the gravitational barrier of the BH and move to infinity. For the gas with a lower velocity may still be accelerated to a high speed to escape the system if the total luminosity of the disk is super-Eddington, which seems always to be satisfied in the cases with outflows considered in this work (see Figure \ref{fig5}). Our detailed estimate of the least power needed to accelerate the gas to infinity indeed shows that it is only a small fraction of the radiation power (luminosity) of the disk (usually around one-tenth of the disk luminosity). This is mainly due to the most gas being driven from the outer region of the disk (see Figure \ref{fig2}), and the gas in the outflows only need to overcome a rather shallow BH gravitational potential barrier.

We note that the model calculation results for the disk structure vary slightly with the value of viscosity parameter $\alpha$, {though the properties of the outflows (power and velocity) vary significantlly with $\alpha$ (see Figure \ref{fig9}).} The density of the disk decreases with increasing $\alpha$, as the radial velocity of the disk is proportional to $\alpha$ for given $\dot{m}$ at $r_{\rm out}$, which leads to the decreases of mass loss rate $\dot{m}_{\rm w}$ in the outflows (see Figure \ref{fig2}), {though the gas with a lower density can therefore be accelerated by the radiation force in the disk to a higher velocity at the disk surface (see Figure \ref{fig9}).} Thus, the accretion rates decrease more rapidly with decreasing radius especially in the outer disk regions for a lower value of $\alpha$ (see Figure \ref{fig2}), while the accretion rate in the inner disk region is almost independent of $\alpha$, which causes slight discrepancies in the continuum spectra with different values of $\alpha$ (cf., Figures \ref{fig6}, \ref{fig8} and \ref{fig9}). Such discrepancies are more evident in the extremely high accretion rate cases, because the removal fraction of the gas into the outflows is less when the accretion rate is extremely high, and a substantial fraction of the gas fed onto the disk is swallowed by the disk, which depends on $\alpha$ more sensitively than the lower accretion rate cases (see Figure \ref{fig2}).

The disk luminosity increases slowly with the accretion rate at the outer radius of the disk (see Figure \ref{fig5}). Most gas in the outflows is driven from the outer region of the disk (Figure \ref{fig2}), which means that the power needed to accelerate the outflows to infinity is less than that for the gas driven from the inner region of the disk. The accretion rates in the inner region are always around $\sim 2\dot{M}_{\rm Edd}$ for the disks with outflows when the accretion rates are as high as $\sim 100\dot{M}_{\rm Edd}$, which indicates a rather similar disk structure in the region near the BH. Indeed, we find that the continuum spectra of the disks saturate at high end frequency, while the situation of the extremely high accretion rate cases is different (see Figure \ref{fig6}). The radiation of the disk seems to saturate in the low and high end of the continuum spectra at around several Eddington luminosities, which is dominantly regulated by the mass loss in the outflows. The saturation luminosity calculated in this work is lower than that of a slim disk, which is roughly consistent with the hydrodynamical numerical simulations \citep*[][]{2005ApJ...628..368O}.

{In the calculations of the vertical structure of the disk, a polytropic index is adopted for simplicity. The index $\gamma=1.5$ is inferred for a vertically hydrostatic radiation pressure dominant accretion disk \citep[][]{1981AcA....31..283P,1988ApJ...332..646A}. For the disk with outflows considered in this work, a polytropic index $\gamma$ is also employed in calculating the vertical structure of the disk. In order to investigate how the results may be affected by the value of $\gamma$, we calculate the structure of the disk and outflows for different values of $\gamma$ while the values of all the other parameters are fixed (see Figure \ref{fig10}). It is found that both the structures of the disks and the outflows are quite insensitive to $\gamma$. It is worth to note that the radiation flux/spectrum of the disk barely changes with $\gamma$. }

It was predicted elegantly that the inflowing rate of the gas onto an accretion disk formed after a TDE varies with time as $\propto t^{-5/3}$ \citep*[][]{1988Natur.333..523R}. This can be translated to a light curve by assuming a constant radiation efficiency, which is widely taken as a criterion for identifying TDEs, however, there is accumulating evidence of TDE light curves significantly deviating from $t^{-5/3}$ \citep*[e.g.,][and the references therein]{2016MNRAS.455.2918H,2018MNRAS.474.3593K}. It is found that almost all observed TDE light curves peaked at around Eddington luminosity, while a substantial fraction of TDEs exhibit flat light curves, i.e., they decline with time more slowly than $t^{-5/3}$ \citep*[e.g.,][]{2016MNRAS.455.2918H,2020ApJ...898..161H}. We apply our model to the observed light curves of the TDEs in Section \ref{results}, and find that the peak luminosity is $\sim 2-3L_{\rm Edd}$, and the radiation of the disk decreases rather slowly with time, which are qualitatively consistent with the major observed features of TDEs (considering the estimate accuracy of BH mass in TDEs is around several times). Our model calculations imply strong winds from the disks in the early stage of the TDEs, which is consistent with the observations and the numerical simulations of TDEs \citep*[][]{2018MNRAS.474.3593K,2018ApJ...859L..20D,2022arXiv220314994H}. About a half of the stellar mass is inflowing to form an accretion disk after a star is disrupted by the tidal force of the BH \citep*[][]{1988Natur.333..523R}. Integrating the light curve of the TDE, one could estimate the total energy radiated from the TDE, and find its radiation efficiency significantly lower than that of a standard thin disk \citep*[e.g.,][]{2016MNRAS.455.2918H,2020ApJ...898..161H}. This feature can be naturally explained in the frame of our model. At the early phase of the TDE, the initial inflowing mass rate is at a hyper-Eddington rate, a substantial fraction of the inflowing gas onto the accretion disk is driven by the radiation force of the disk into the outflows, and the accretion rate in the inner disk region is significantly reduced, which leads to the disk radiating at $\sim 2-3L_{\rm Edd}$. The radiation efficiency calculated with the inflowing gas rate at the outer edge of the disk is therefore much lower than that of a standard thin disk. The detailed comparison with the TDE observations is beyond the scope of this work, which will be reported in our future work.

The discovery of super-massive BHs at high-$z$ provides important clues on the BH growth \citep*[e.g.,][and the references therein]{2011Natur.474..616M,2015Natur.518..512W,2016ApJ...833..222J,2018ApJ...869L...9W}. Our results show that strong outflows are always associated with super-Eddington accretion disks, and only a small fraction of the inflowing gas is swallowed by the BHs, which implies that the rapid growth of super-massive BHs through supercritical accretion is rather inefficient, unless the inflowing mass rate onto the disk at extremely high rates ($\ga$ several hundred Eddington-scaled rates, see Figure \ref{fig2}). The outflows with a high mass loss rate driven by a super Eddington accretion disk must play crucial roles in the host galaxy, namely, mechanical feedback \citep*[e.g.,][]{2010ApJ...722..642O,2010MNRAS.409..985D,2011ApJ...737...26N,2012ApJ...754..125C,2018ApJ...857..121Y,2019ApJ...877...47Q}. The results of this work can be taken as an ingredient in the mechanical feedback models.

\acknowledgments
We are grateful to the referee  for his/her constructive suggestions/comments.
We thank Bei You, Qingwen Wu, and Tinggui Wang for helpful discussion/comments. This work is supported by the NSFC (11773050, 11833007, and 12073023), and the science
research grants from the China Manned Space Project with NO. CMS-CSST-
2021-A06, and a startup grant from Zhejiang University.

{}

\end{document}